\documentclass[a4paper,11pt]{article}
\usepackage[a4paper, portrait, margin=1in]{geometry}
\usepackage{authblk}
\usepackage{hyperref}
\usepackage{wrapfig}
\usepackage{graphicx}
\usepackage{amsfonts}
\usepackage[ruled,vlined]{algorithm2e}
\newsavebox{\tempfig}
\usepackage{graphicx, floatrow}
\usepackage[utf8]{inputenc}
\usepackage[english]{babel}
\usepackage{caption}
\usepackage{subcaption}
\usepackage{amsthm}
\usepackage{amsmath}
\usepackage{amssymb}
\usepackage{physics}
\usepackage[utf8]{inputenc}
\usepackage[english]{babel}
\usepackage[table,usenames,dvipsnames]{xcolor}
 \usepackage{textcomp}
 \usepackage[table]{xcolor}
 \usepackage{multirow}
 \usepackage{mathrsfs,amsmath}
\usepackage[sort&compress,square,numbers]{natbib}
\usepackage{arydshln}

\usepackage{tikz}
\usetikzlibrary{plotmarks}
\usepackage{etoolbox}
\newcommand{\be}{\begin{equation}}
\newcommand{\ee}{\end{equation}}
\usepackage{caption}
\usepackage{graphicx}
\usepackage{subcaption}
\usepackage{etoolbox}
\DeclareRobustCommand\full  {\tikz[baseline=-0.6ex]\draw[blue,thick] (0,0)--(0.5,0);}
\DeclareRobustCommand\fullgreen{\tikz[baseline=-0.6ex]\draw[black!40!green,thick] (0,0)--(0.5,0);}
\DeclareRobustCommand\fullred{\tikz[baseline=-0.6ex]\draw[red!40!red,thick] (0,0)--(0.5,0);}
\DeclareRobustCommand\fullblack 
{\tikz[baseline=-0.6ex]\draw[black,thick] (0,0)--(0.5,0);}
\DeclareRobustCommand\fullblue 
{\tikz[baseline=-0.6ex]\draw[blue,thick] (0,0)--(0.5,0);}
\DeclareRobustCommand\fullgray  {\tikz[baseline=-0.6ex]\draw[gray,thick] (0,0)--(0.5,0);}
\DeclareRobustCommand\dotted{\tikz[baseline=-0.6ex]\draw[thick,dotted] (0,0)--(0.54,0);}
\DeclareRobustCommand\dashedgreen{\tikz[baseline=-0.6ex]\draw[black!50!green,thick,dashed] (0,0)--(0.54,0);}
\DeclareRobustCommand\dashedred{\tikz[baseline=-0.6ex]\draw[red,thick,dashed] (0,0)--(0.54,0);}
\DeclareRobustCommand\dashedblack{\tikz[baseline=-0.6ex]\draw[black,thick,dashed] (0,0)--(0.54,0);}
\DeclareRobustCommand\dashedblue{\tikz[baseline=-0.6ex]\draw[blue,thick,dashed] (0,0)--(0.54,0);}

\DeclareRobustCommand\chainblack {\tikz[baseline=-0.6ex]\draw[black, thick,dash dot ] (0,0)--(0.5,0);}
\DeclareRobustCommand\chainred {\tikz[baseline=-0.6ex]\draw[red, thick,dash dot ] (0,0)--(0.5,0);}
\newrobustcmd*{\mycircle}[1]{\tikz{\filldraw[draw=#1,fill=#1] (0,0) circle [radius=0.05cm];}}
\newrobustcmd*{\mytriangle}[1]{\tikz{\filldraw[draw=#1,fill=#1] (0,0)--(0.2cm,0) -- (0.1cm,0.2cm);}}

\title{State estimation in minimal turbulent channel flow: \\ 
A comparative study of 4DVar and PINN}
\author{ 
 Yifan Du, 
 Mengze Wang, 
  Tamer A. Zaki\thanks{Email address for correspondence: t.zaki@jhu.edu}
}
\affil{\small Department of Mechanical Engineering, Johns Hopkins Univeristy, Baltimore, MD 21218, USA}

\date{}

\begin{document}
\maketitle

\begin{abstract}
The state of turbulent, minimal-channel flow is estimated from spatio-temporal sparse observations of the velocity, using both a physics-informed neural network (PINN) and adjoint-variational data assimilation (4DVar). The performance of PINN is assessed against the benchmark results from 4DVar. 
% The full resolution spatio-temporal field is estimated from sparse velocity data generated from a reference direct numerical simulation. 
The PINN is efficient to implement, takes advantage of automatic differentiation to evaluate the governing equations, and does not require the development of an adjoint model.  
% and takes advantage of GPU parallelis for the forward and backward propagation of network are massively parallelized. 
% In addition, the flow evolution is expressed in terms of the network parameters which have a smaller dimension than the initial condition of the forward simulation. 
In addition, the flow evolution is expressed in terms of the network parameters which have a far smaller dimension than the predicted trajectory in state space or even just the initial condition of the flow. 
Provided adequate observations, network architecture and training, the PINN can yield satisfactory estimates of the the flow field, both for the missing velocity data and the entirely unobserved pressure field.  
However, accuracy depends on the network architecture, and the dependence is not known a priori.  
In comparison to 4DVar estimation which becomes progressively more accurate over the observation horizon, the PINN predictions are generally less accurate and maintain the same level of errors throughout the assimilation time window. 
Another notable distinction is the capacity to accurately forecast the flow evolution: while the 4DVar prediction depart from the true flow state gradually and according to the Lyapunov exponent, the PINN is entirely inaccurate immediately beyond the training time horizon unless re-trained.  
Most importantly, while 4DVar satisfies the discrete form of the governing equations point-wise to machine precision, in PINN the equations are only satisfied in an $L^2$ sense.
% The architecture and training of PINN does not reflect causality that are inherent in the governing equations of fluid dynamics problems, thus the reconstruction error from PINN is uniform on time. 
\end{abstract}

%================================
%       INTRODUCTION
%================================
\section{Introduction}

Simulations and experiments are complementary approaches in the study of wall turbulence.  
The respective benefits of both methods can be combined synergistically, when simulations are infused with experimental measurements.  
The assimilation of the observations endows the simulations with a unique level of fidelity, because their predictions are transformed from being a generic trajectory in state space to the exact one realized in the experiment. 
In addition, the simulations compute the flow state at full resolution, far beyond the original measurements, and can also directly evaluate quantities that are not measured such as pressure. 
Data assimilation, or flow estimation from limited observations, can be performed using a variety of techniques including filtering \citep{Evensen1994EnKF,Suzuki2017}, smoothing using adjoint \citep{zaki2021prf} and ensemble variational algorithms \citep{Buchta2021envar,Buchta2022} and recently using physics-informed machine learning \citep{raissi2019physics}. In this study, we consider sparse observations from an independent simulation of minimal turbulent channel flow, and perform state estimation using both adjoint-variational data assimilation (4DVar) and physics informed neural network (PINN).   Using the benchmark results from 4DVar, we provide a detailed analysis on the accuracy of the estimation using PINN.

Data assimilation methods for flow reconstruction can be classified based on a number of criteria, perhaps most important among them in fundamental studies of turbulence is whether the governing equations are enforced. If the governing equations are not enforced, a statistical or dynamical model is adopted instead. For example, linear stochastic estimation \citep{Adrian1988LSE,Jimenez2019LSE} uses prior knowledge of two-point correlations to estimate the full velocity field from observations; 
Extended Kalman filter provides an optimal update to the state by combining a prediction from the linearized equations and the observations  \citep{Bewley_part2,Suzuki2012}, while ensemble Kalman filter advances an ensemble of solutions and optimally weighs them based on the observations to estimate the new state \citep{colburn2011state}.
Some recent machine-learning algorithms also belong in this class, when neural networks learn a mapping from limited measurements to full flow field, based on training data from simulations and without direct enforcement of the Navier-Stokes equations.  In such instances, there is no guarantee that predictions, especially outside the training space, satisfy the governing equations.  
For example, \citet{fukami2019super} trained a convolutional neural network to map from coarse-grained to full-resolution velocity fields based on training data from two-dimensional homogeneous turbulence;
In \citet{gundersen2021semi}, a semi-conditional variational auto-encoder was developed to perform flow reconstruction from sparse measurements in a probabilistic framework, which predicts the full flow field as well as it uncertainty.  In these methods, the predicted fields do not satisfy the physical governing equations, and some of the estimated flow structures could be the outcome of generalization errors of the surrogate model instead of physical ones governed by the Navier-Stokes equations.

The second class of methods aims to predict a trajectory of the flow in state space, that both satisfies the governing equations and optimally reproduces the measurements.  In this class, four-dimensional adjoint variational data assimilation \citep{Dimet1986_4dvar,Li2020}, or 4DVar, casts the problem as an optimization constrained by the governing equations. 
%   The loss function is defined in terms of the discrepancy between the available measurements and their estimates from the simulation. 
Starting from an estimate of the unknown flow state, a forward simulation produces the full flow trajectory over the time horizon where measurements are available. The disparity between the measurements and their estimates from the simulation defines the loss function, and also features in the adjoint equations which are solved backward in time. A complete forward-adjoint loop yields the gradient of the loss with respect to the unknown flow state, which can be adopted to improve the estimate of the state. Since the governing equations are nonlinear, the procedure is repeated till convergence, whereby the optimal flow state is identified and accurately reproduces the entire history of observations.  
The method has been adopted in a wide range of applications, including prediction of scalar sources from remote measurements \citep{Wang_hasegawa_zaki_2019}, estimation of transitional and turbulent Taylor-Couette flows from limited observations \citep{wang2019discrete}, and estimation of turbulence in channel flow \citep{bewley2004skin,Foures2014,wang2021state,wang2022hessian}.
%   \citet{bewley2004skin} used adjoint method to reconstruct the velocity field in the channel from noisy wall measurements, at a modest friction Reynolds number of $Re_\tau = 100$. \citet{wang2021state} estimated channel using spatio-temporal distributed measurements at varying resolution, and also examined flow estimation from wall data.  For the latter configuration, they considered a range of Reynolds numbers.  Due to the accurate and efficient calculation of gradients from adjoint simulation and exact satisfaction of governing equation for flow fields, the adjoint methods outperform most of similar methods for flow reconstructions. 
An important property of 4DVar is that the computational cost of evaluating the gradient of the loss function is one forward-adjoint loop, independent of the size of the control vector being optimized.  This efficiency presents an advantage compared to other optimization algorithms that only adopt the forward model. For example, in ensemble-variational (EnVar) approaches \citep{mons2019kriging,mons2021les} the gradient of the loss function is evaluated from an ensemble of forward solutions whose size, and hence the associated computational cost, are proportional to the dimension of the control vector.  \citet{mons2016reconstruction} performed a comparison of 4DVar, ensemble-variational (EnVar) method and ensemble Kalman filter, and concluded that 4DVar is the most accurate.  In this study, the adjoint reconstruction of flow field will be adopted as the benchmark for evaluating the performance of the physics-informed, machine-learning approach.

Recent innovation in machine learning has presented new opportunities for data assimilation and flow estimation \citep{mao2020physics,lou2021physics,mao2021jcp,cai2021jcp,PCDL2021DeepONet}.  
Our primary focus will be on physics-informed neural networks (PINNs).  Similar to the adjoint-variational approach, flow estimation using PINNs is formulated as a minimization problem.  The network inputs are the spatial and temporal coordinates and its outputs are the flow variables at the corresponding coordinates.  The loss function for training the PINN is comprised of different parts: (a) The first part is due to the mismatch between the network predictions and the available flow measurements; (b) The second part is in terms of the residuals of the governing equations and other constraints, e.g.\,the boundary conditions. The minimization of such loss function leads to a neural network which predicts correct observation values and satisfies the governing equations at the training spatio-temporal locations, in a weak sense. 
\citet{raissi2019physics} introduced the notion of PINNs for solving partial differential equations (PDEs). The ease of implementation led to a number of applications with different types of PDEs and flow conditions, for example \citet{mao2020physics} considered forward and inverse problems in high-speed flows; \citet{lou2021physics} examined rarified flow with Bolzmann-BGK formulation; \citet{mowlavi2021optimal} adopted the PINN framework for optimal control and validated the approach against adjoint-based nonlinear control;  and \citet{yang2021b} developed the Bayesian PINN methodology for inverse PDE problems with noisy data. 

In contrast to adjoint methods which enforce the governing equations exactly and preserve causality in the predicted state-space trajectory, PINNs methods use the residual of the governing equation as a penalty in the optimization problem.  As such, the PINN-estimated flow only satisfy the governing equations in $L^2$ sense.  The advantage of PINN, however, is the ease of implementation for any set of forward evolution equations, and without the need for an adjoint model.  
% The adjoint methods require a development of adjoint simulation tool along with the forward simulation code for a certain type of governing equation, while the PINNs methods are easy to implement, and could easily be adopted to different types of equations. 
We will provide a practical guide for application of PINN in estimation of turbulent flows from limited observations, and discuss various properties of these networks.

In this study, we perform estimation of turbulent flow in a minimal channel configuration, from sparse velocity measurements using 4DVar and PINNs.  The results from the adjoint-variational approach provide the benchmark for evaluating the accuracy of the PINN predictions.  In \S\ref{sec:method}, we formulate the data assimilation problem and present the details of the 4DVar and PINNs algorithms.  The results are presented and discussed in \S\ref{sec:results}, followed by a summary of the main conclusions in \S\ref{sec:conclusion}. 

%================================
%       METHODOLOGY
%================================

\section{Methodology}
\label{sec:method}

%-----------------------------
%       State estimation
%-----------------------------
\subsection{The state-estimation problem}
\label{sec:forward}

\begin{figure}
    \centering
    \includegraphics[width=1.0\textwidth]{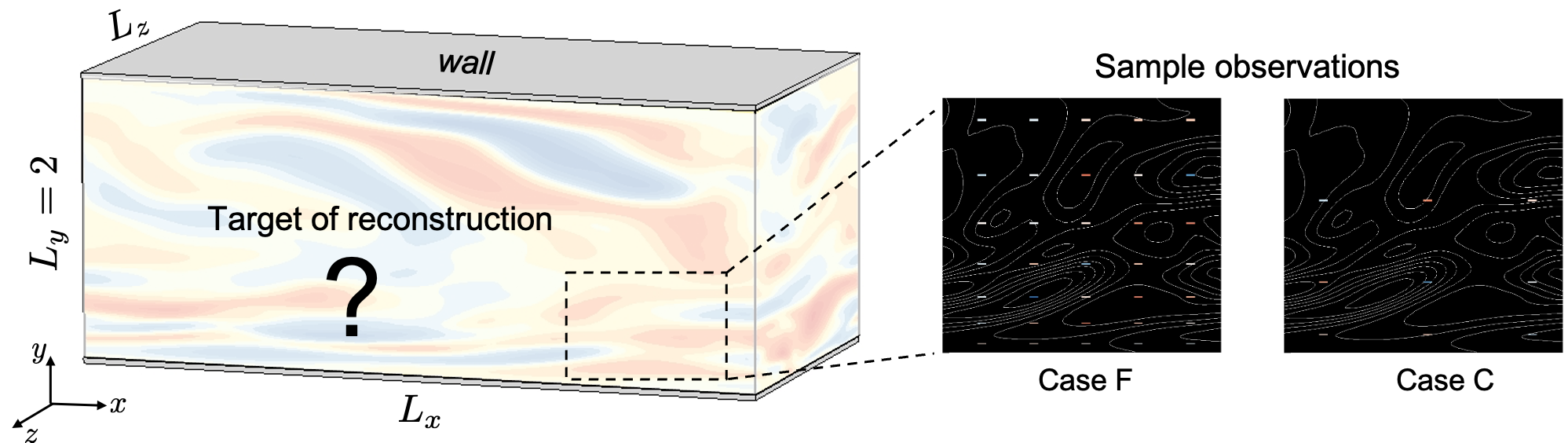}
    \caption{Schematic of the reference simulation of turbulent channel flow and sample observations. Right two panels show a zoomed-in region of the flow, and contrast the fine and coarse observations (color rectangle) to the background line contours of the full, unknown flow field.}
    \label{fig:schematics observation}
\end{figure}

In this section, we formulate the data-assimilation problem, where we attempt to estimate the initial state of channel-flow turbulence from sparse measurements. 
The channel half height $h^{*}$ and bulk flow velocity $U^{*}$ are adopted as the reference scales, where the star denotes dimensional quantities.  The flow domain, shown in figure \ref{fig:schematics observation}, is a minimal rectangular flow unit, 
%$\Omega = \{(x,y,z)\in [0,1.72]\times[0,2]\times[0,0.87]\}$
$\Omega = \{(x,y,z)\in [0,L_x]\times[0,2]\times[0,L_z]\}$. 
The fluid motion is bounded by two fixed walls at $y = 0$ and $y = 2$, is periodic on both the horizontal $x$ and $z$ directions, and is governed by the non-dimensional incompressible Navier-Stokes equations, 
\begin{eqnarray}\label{eq:NS}
\label{eq:cont_div}
    \nabla \cdot \boldsymbol{u} &=& 0  \\
	\label{eq:cont_mom}
	\frac{\partial \boldsymbol{u} }{\partial t} +  \boldsymbol{u} \cdot \nabla \boldsymbol{u} + \nabla p - \frac{1}{\mathrm{Re}} \nabla^2 \boldsymbol{u} &=& 0
\end{eqnarray}
where $\boldsymbol{u}=(u, v, w)$ is the velocity, and $p$ is the pressure. 
The bulk Reynolds number is $\mathrm{Re} \equiv U^*h^*/\nu^*$, where $\nu^*$ is the fluid kinematic viscosity.
The velocity satisfies no slip at the walls, and is periodic on $x$ and $z$ boundaries.  
The flow is statistically stationary, and is sustained by a known constant mean pressure gradient in the $x$ streamwise direction. The total pressure is decomposed into $p = P + p'$, where $P$ is the mean pressure and $p'$ is the fluctuating pressure that is periodic in $x$ and $z$ directions. 
The velocity and pressure boundary conditions can be expressed in operator form as, 
\begin{align}\label{eq: BC forward}
    \mathscr{B}[\boldsymbol{u},p']
         =\boldsymbol{0}.
\end{align}

The friction velocity is defined as $u_{\tau}^* \equiv \sqrt{\overline{\tau_w^*}/\rho^*}$, where $\overline{\tau_w^*}$ is the mean shear stress on the walls, and the friction Reynolds number is given by $\mathrm{Re}_{\tau} \equiv u_{\tau}^*h^*/\nu^*$. 
%   \del{Two Reynolds numbers are considered, $\mathrm{Re}_{\tau}=\{180,392\}$; the majority of the discussion will be focused on $\mathrm{Re}_{\tau}=180$ which is highlighted in table \ref{table:channel setup}.}
The discussion in the main text will be focused on $\mathrm{Re}_{\tau}=180$, and the effect of a higher Reynolds number ($\mathrm{Re}_{\tau}=392$) will be discussed in Appendix \ref{sec:appendix B}.
When viscous scaling is adopted for non-dimensionalization, quantities will be marked by superscript `+'.  For example, the domain sizes in viscous units, $L^{+} = L^* u_{\tau}^*/\nu^*$, are reported in table \ref{table:channel setup}.  The present values are motivated by previous studies of the minimal flow unit, which determined the domain sizes required to sustain wall turbulence \citep{jimenez1991minimal,podvin1998low,flores2010hierarchy}. Specifically, the statistics of minimal channel in the near-wall region agree with those in large channels when the spanwise and streamwise sizes of the former are larger then $100$ wall units \citep{jimenez1991minimal}.   

	\begin{table}
		\centering
		\begin{tabular}{c c c c c}
			Domain Size  & \ Grid points \ &  Grid resolution  & Reynolds number\\
			
			\begin{tabular}{c c c}
				\hline 
				$L_x^{+}$ & $L_y^{+}$ & $L_z^{+}$ \\ \hline 
			%\rowcolor{blue!20}	
			314   &   360   &  157   \\
			% Re_tau = 392 case is moved to appendix. 
		    %\tz{314}   &  \tz{780}   &  \tz{157}
			\end{tabular} & 
			\begin{tabular}{c c c}
				\hline
				$N_x$ & $N_y$ & $N_z$ \\ \hline 
			%\rowcolor{blue!20}	
			65     & 385    & 65  \\
			%\tz{64}    & \tz{896}    & \tz{64}
			\end{tabular} & 
			\begin{tabular}{c c c c}
				\hline
				$\Delta x^+$ & $\Delta y^+_{min}$ & $\Delta y^+_{max}$ & $\Delta z^+$ \\ \hline 
				%\rowcolor{blue!20}	
				4.90             &   0.40           &   1.40                     &  2.45  \\
	            %\tz{4.90}            &   \tz{0.30}           &   \tz{1.30}               &  \tz{2.45 }\\
			\end{tabular} &
			\begin{tabular}{c c}
				\hline
				$\mathrm{Re}$ & $\mathrm{Re}_{\tau}$  \\ \hline 
			%\rowcolor{blue!20}	
			2800            &   180          \\
		    %\tz{6875}            &   \tz{392}
			\end{tabular}\\
		\end{tabular}
		\caption{Domain sizes and grid resolutions of reference DNS.}
		\label{table:channel setup}
	\end{table}

The flow estimation problem relies on the availability of some measurements, or observations.  In the present study, the measurements are spatio-temporal sparse velocity data, without any knowledge of the pressure field.  Our goal is to reconstruct the full spatio-temporal velocity and pressure fields. The objective is therefore to identify the initial flow state that, when evolved using the governing equations (\ref{eq:NS}), reproduces all the available measurements. The estimation of the initial condition can therefore be cast as the constrained optimization problem: 
\be\label{eq:NS optimization}
    \begin{split}
    \min_{\boldsymbol{u},p}& \quad \, \mathcal{J}_m = \frac{1}{2}\left\Vert \mathcal{M}\left(\boldsymbol{u}\right) - \boldsymbol{m}\right \Vert^2, \\
    \text{subject to}& \quad \, \mathscr{N}[\boldsymbol{u},p] = 0, \quad \, \, \mathscr{B}[\boldsymbol{u},p] = 0, 
    \end{split}
\ee
where $\mathcal{J}_m$ is the measurement loss, $\boldsymbol{m}$ represents sparse measurement velocity data from experiments or reference numerical simulations, and $\mathcal{M}$ is the measurement operator that maps the continuous spatio-temporal solution $\boldsymbol{u}$ to its values at the same discrete measurement locations. The operator $\mathscr{N}$ maps functions $\boldsymbol{u}$ and $p$ to the residuals of equations (\ref{eq:NS}). When the equations are strongly enforced, the residuals are exactly zero, and the instantaneous forward velocity fields are fully determined by the initial condition, $\boldsymbol{u} = \boldsymbol{u}(\boldsymbol{u_0})$. As such, the loss function is fully determined by the initial condition, $\mathcal{J}_m = \mathcal{J}_m(\boldsymbol{u}(\boldsymbol{u_0}))$, and its derivative is denoted as $\mathcal{D}\mathcal{J}_m / \mathcal{D}\boldsymbol{u_0}$.

In order to ensure accuracy and have full control of the resolution of measurements, we acquired them from an independent reference direct numerical simulation (DNS).  Once the measurements are collected, the true flow is hidden throughout the data assimilation procedure, and is only revisited to quantify the accuracy of the estimated initial conditions. 
The algorithm for the reference DNS was used in a number of earlier studies of wall turbulence, and has been extensively validated \citep{Jelly2014,you_zaki_2019}.  It adopts a fractional-step approach with a local volume flux formulation on a staggered grid \citep{Rosenfeld1991}. The advection terms are discretized by the Adams-Bashforth scheme, and the Crank-Nicolson scheme is adopted for the diffusion terms. The pressure Poisson equation is solved using Fourier transforms in the periodic directions and tri-diagonal inversion in the wall-normal direction.  
The reference solution $\boldsymbol{u}_r$ is then sub-sampled in both space and time. The sampling gaps are expressed in terms of the number of discretization points $(\Delta M_x, \Delta M_y , \Delta M_z, \Delta M_t)$ in table (\ref{table:obs}). Previous studies have demonstrated that the maximum observation gap that allows accurate reconstruction of the full velocity field corresponds to the correlation lengthscale of the turbulence \citep{wang2021state,wang2022synch}, which is measured by the Taylor microscale:
\be
    \Lambda^{\xi}_{r} = \left(-\frac{1}{2}\frac{\mathrm{d}^2R_{\xi}(r)}{\mathrm{d}r^2}\right)^{-1/2}
\ee
where $R_{\xi}(r)$ is the two-point correlation function of $\xi$ velocity component in the $r$ coordinate direction. 

Two resolutions of the measurements will be considered, the finer of which is denoted case `F' and the coarse as case `C' (see table \ref{table:obs} and also figure \ref{fig:schematics observation}). 
The gap between observations in the former case is one Taylor microscale, and twice that scale in case `C'. 
In all cases, observations are available within a time horizon $0 \le t \le T$, where in viscous units $T^+ = 50$. 
This duration is approximately one Lyapunov timescale ($\tau_{\sigma}^+ \approx 48$ for $\mathrm{Re}_\tau = 180$), which has weak Reynolds-number dependence in the range considered \citep{Nikitin2018,wang2022hessian}. 
In all cases, the spatio-temporal resolution of the observations is not sufficient to achieve synchronization of the estimated state to the true flow \citep{wang2022synch}. As a result, the 4DVar and PINN are not expected to perfectly reproduce the true flow, and we can compare their prediction accuracy.

\begin{table}
	\centering
		\begin{tabular}{c c c c c c c c c c c }
			 Case & $\Delta M_x $ & $\Delta M_y$ & $\Delta M_z$ & $\Delta M_t$ & $\Delta x_{m}^+$   & $\Delta y^+_{m}$  & $\Delta z_{m}^+$ & $\Delta t^+_{m}$  & $T^+$ \\
			\hline 
			 %R100 & 8 & 8 & 8 & 8  & 39.2 & [2.4, 9.6] & 19.6 & 2.2 &  49.26 \\
			 F  & 6 & 24 & 4 & 32  & 32 & [9.6, 33.2] & 10 & 2.27 &  50.0 \\
			 C  & 12 & 48 & 8 & 64  & 64 & [19.2, 66.4] & 20 & 4.54 &  50.0 \\
			 %\tz{F-392} & \tz{392} & 6 & 32 & 4 & 32  & 32 & [9.6, 41.6] & 10 & 2.17 &  50.0 \\
	\end{tabular}
	\caption{Parameters of observations extracted from the reference simulation. The sampling rate relative to the DNS is $\Delta M$, with a subscript that denotes the spatial or temporal dimension.  The spatio-temporal resolution of the measurements (subscript `$m$') and the time horizon $T$ are also provided, in viscous units (superscript `+'). }
	\label{table:obs}
\end{table}
	
% \begin{figure}\label{fig:schematics}
%     \centering
%     \begin{subfigure}{.55\linewidth}
%     \centering
%     \includegraphics[width=1.0\textwidth]{schematics/Adjoint_schematics_v3.png}
%     \caption{Adjoint-variational approach (4DVar)}
%     \label{fig:schematics adjoint}
%     \end{subfigure}%
%     %
%     \begin{subfigure}{.45\linewidth}\label{fig:pinn}
%     \centering
%     \includegraphics[width=1.0\textwidth]{schematics/PINN_schematics.png}
%     \caption{Physics-informed Neural Network (PINN)}
%     \label{fig:schematics pinn}
%     \end{subfigure}
%     \caption{ Schematics of (a) adjoint-variational approach and (b) PINN. (a) Forward evolution of estimated initial condition yields discrepancies with available measurements.  The adjoint evolution provides the gradient of the Lagragian with respect to the initial condition which is update.  (b) Training of PINNs, with both space and time as inputs, and velocity and pressure as outputs. The loss includes observations, boundary conditions and equations.}
%     \label{fig:Re100}
% \end{figure}

%-----------------------------
%       Adjoint
%-----------------------------
\subsection{Adjoint-variational state estimation}
\label{sec:4DVaralgorithm}

\begin{figure}
    \centering
    \includegraphics[width=1.0\textwidth]{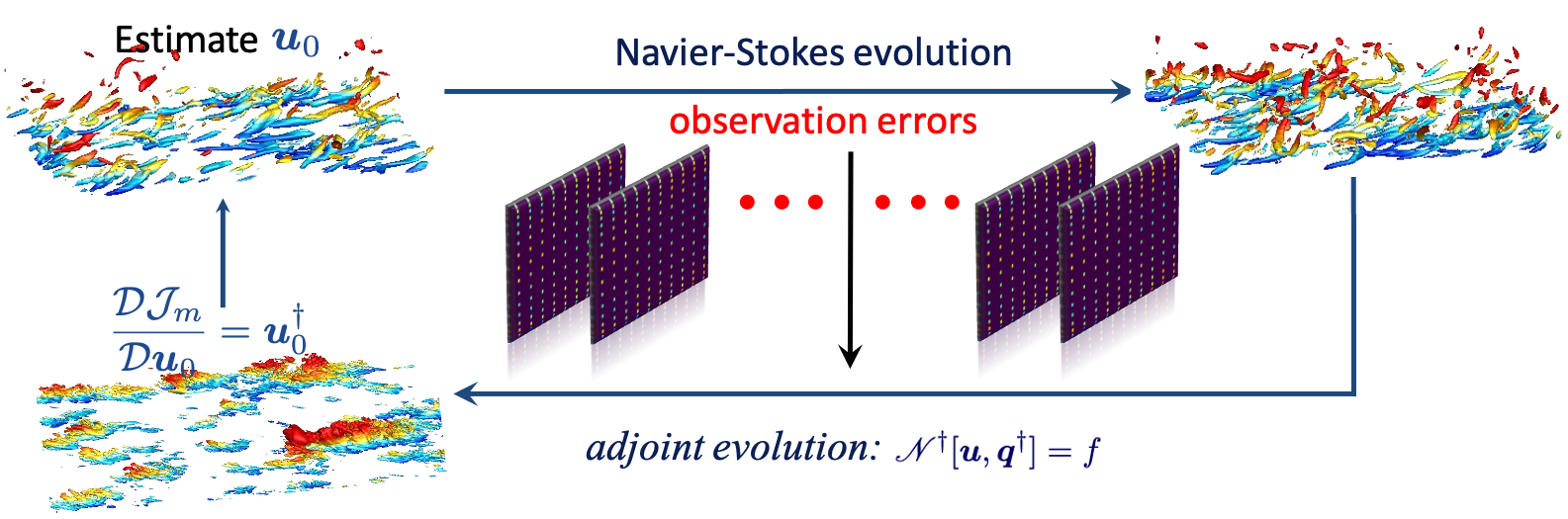}
    \caption{Schematic representation of 4DVar. Forward evolution of an estimated initial condition $\boldsymbol{u}_0$ does not reproduce the available observations. The observations errors force the adjoint equations which are evolved backward in time. The resulting adjoint field is the gradient of the loss function with respect to the initial condition, and is used to update $\boldsymbol{u}_0$. This procedure is repeated until convergence.}
    \label{fig:schematics adjoint}
\end{figure}

The optimization problem (\ref{eq:NS optimization}) can be reformulated in terms of the Lagrangian, 
\be
    \mathcal{L} = \mathcal{J}_m + \left<\boldsymbol{q}^{\dag},\mathscr{N}[\boldsymbol{u},p]\right> + \left<\boldsymbol{\beta}^{\dag},\mathscr{B}[\boldsymbol{u},p]\right>
\ee
% \textcolor{blue!40!black!}{\be
%     \mathscr{N}^{\dag}[\boldsymbol{u},\boldsymbol{q}^{\dag}] = f
% \ee}
where the forward state is $\boldsymbol{q} \equiv ( u, v, w, p)$ and its adjoint is $\boldsymbol{q}^{\dag} = (p^{\dag}, u^{\dag}, v^{\dag}, w^{\dag})$.  For convenience, we also introduce forward and adjoint variables, $\boldsymbol{\beta}$ and $\boldsymbol{\beta}^{\dag}$, for enforcing the boundary conditions.  The first order optimality condition yields the following equalities: 
\be
    \begin{split}
    \frac{\mathscr{D}\mathcal{L}}{\mathscr{D}\boldsymbol{q}^{\dag}} &= 0, \quad \, \, \frac{\mathscr{D}\mathcal{L}}{\mathscr{D}\boldsymbol{\beta}^{\dag}} = 0, \\
    \frac{\mathscr{D}\mathcal{L}}{\mathscr{D}\boldsymbol{q}} &= 0,\quad \, \, \frac{\mathscr{D}\mathcal{L}}{\mathscr{D}\boldsymbol{\beta}} = 0,
    \end{split}
\ee
where $\mathscr{D}$ denotes the functional derivative. The first two conditions are gradients of the Lagrangian with respect to the adjoint variables, which lead to the Navier-Stokes equations (\ref{eq:NS}) and boundary conditions (\ref{eq: BC forward}). The last two conditions are gradients of the Lagrangian with respect to the primal flow variables and primal boundary values, which lead to the adjoint equations and associated boundary conditions: 
\be\label{eq:adjoint}
    \begin{split}
    \nabla \cdot \boldsymbol{u}^{\dagger} & =\frac{\mathscr{D} \mathcal{J}_m}{\mathscr{D} p} \\
%    \frac{\partial \boldsymbol{u}^{\dagger}}{\partial t^{\dagger}}-\boldsymbol{u} \cdot \nabla \boldsymbol{u}^{\dagger}+(\nabla \boldsymbol{u}) \cdot \boldsymbol{u}^{\dagger}&=\nabla p^{\dagger}+\frac{1}{\operatorname{Re}} \nabla^{2} \boldsymbol{u}^{\dagger}-\frac{\mathscr{D} \mathcal{J}_m}{\mathscr{D} \boldsymbol{u}}, 
    \frac{\partial \boldsymbol{u}^{\dagger}}{\partial t^{\dagger}}-\boldsymbol{u} \cdot \nabla \boldsymbol{u}^{\dagger}+(\nabla \boldsymbol{u}) \cdot \boldsymbol{u}^{\dagger}&-\nabla p^{\dagger}-\frac{1}{\operatorname{Re}} \nabla^{2} \boldsymbol{u}^{\dagger} = -\frac{\mathscr{D} \mathcal{J}_m}{\mathscr{D} \boldsymbol{u}}, 
    \end{split}
\ee
\begin{align}\label{eq: BC adjoint}
     \mathscr{B}^{\dag}[\boldsymbol{u}^{\dag},p^{\dag}] =\boldsymbol{0}.
\end{align}
%   \del{The adjoint equations evolve in reverse time, $t^{\dagger} = T - t$, and feature the forward velocity field $\boldsymbol{u}$ which must therefore be stored over the simulation horizon.}
The adjoint equations (\ref{eq:adjoint}) can be expressed in operator form as $\mathscr{N}^{\dag}[\boldsymbol{u},\boldsymbol{q}^{\dag}] = f$.  They feature the forward velocity field $\boldsymbol{u}$ that must be stored over the simulation horizon.  The forcing term is $f = [\mathscr{D}\mathcal{J}_m / \mathscr{D}p, -\mathscr{D}\mathcal{J}_m / \mathscr{D}\boldsymbol{u}]$, where the derivatives are derived analytically by assuming that the forward fields at different times are independent.  This forcing drives the adjoint solution which is evolved in reverse time, $t^{\dag} = T - t$.
The operator $\mathscr{B}^{\dag}$ describes the boundary conditions of the adjoint variables: The adjoint velocities $(u^{\dag}, v^{\dag}, w^{\dag})$ are zero at the walls and periodic on $x$ and $z$ boundaries; The adjoint pressure $p^{\dag}$ satisfies $\partial p^{\dag} / \partial n = 0$ at the walls, and is periodic on $x$ and $z$ boundaries. 
At the end of adjoint marching, the gradient of loss function with respect to the initial condition is given by the adjoint field at the initial time: 
\be \label{eq:grad}
  \frac{\mathcal{D}\mathcal{J}_m}{\mathcal{D} \boldsymbol{u}_0} = \boldsymbol{u}^{\dag}_0. 
\ee
Note that this gradient is different from the forcing terms in (\ref{eq:adjoint}). The former represents the sensitivity of the loss function to the initial state when the Navier-Stokes equations are strongly enforced, whereas the latter is analytically derived by assuming that the forward fields at different times are independent.
%   This gradient can then be adopted to improve an estimate of the initial flow state. 

Starting from an estimate of the initial condition $\boldsymbol{u}_{0}$, the forward evolution of the flow is simulated by solving equation (\ref{eq:NS}) from $t=0$ to $T$. 
Our first estimate of $\boldsymbol{u}_{0}$ is simply an interpolation of the sparse velocity observations.
During the forward evolution, the discrepancy between the predicted and available measurements are evaluated.  At the end of the observation horizon, the adjoint equations are marched back from $t=T$ (or $t^\dagger=0$) to $t=0$ (or $t^\dagger=T$).  At the end of the adjoint evolution, $\boldsymbol{u}_{0}^{\dag}$ provides the gradient of the loss function with respect to the initial condition (equation \ref{eq:grad}), which is adopted to update our estimate of the initial condition and minimize $\mathcal{J}_m$.   In this study, this update is performed using the limited-memory Broyden-Fletcher-Goldfarb-Shanno (L-BFGS) method \citep{LBFGS}.  The overall procedure of adjoint variational data assimilation is summarized in Algorithm \ref{alg:adjoint}, and a schematic is provided in figure \ref{fig:schematics adjoint}.
The optimization algorithm is performed for one-hundred iterations, or forward-adjoint loops, and hence a fixed computational cost for a given simulation grid.

	\begin{algorithm}[h]
		\SetAlgoLined
		\textbf{Step 1}: Forward model\;
		\Indp
		\textbullet~Start with an estimate of the initial condition $\boldsymbol{u}_{0}$. 
		The first estimate is interpolation of the sparse velocity observations\;
		\textbullet~Advance $\boldsymbol{u}_{0}$ using the forward equations (\ref{eq:NS}) from $t=0$ to $t=T$ and store the instantaneous forward velocity fields at every time step\;
		\textbullet~Calculate the loss function (\ref{eq:NS optimization})\;
		\Indm
		\textbf{Step 2}: Adjoint model\;
		\Indp
		\textbullet~March the adjoint equations (\ref{eq:adjoint}) from $t^{\dag} = 0$ (or $t=T$) to $t^{\dag}=T$ (or $t=0$)\;
		\textbullet~Obtain the gradient of the loss function at the initial time (equation \ref{eq:grad})\;
		\Indm
		\textbf{Step 3}: Update the estimated initial state\;
		\Indp
		\textbullet~Find an improved estimate of the initial state using the L-BFGS algorithm\; 
		\textbullet~Repeat steps 1-3 to convergence or for a prescribed number of iterations. 
		\caption{Adjoint-variational data assimilation.}
		\label{alg:adjoint}
	\end{algorithm}

%-----------------------------
%       PINN
%-----------------------------
\subsection{PINN state estimation}
\label{sec:PINNalgorithm}

Physics-informed neural network (PINN) is a versatile framework to solve partial differential equations (PDE) and assimilate available measurements. In this section, the methodology is described for estimation of channel-flow turbulence. We start by casting the problem (\ref{eq:NS optimization}) as a weakly constrained optimization,  
\be\label{eq:WC optimization}
    \min_{\boldsymbol{u},p} \mathcal{J}\left(\boldsymbol{u},p\right), \quad \mathcal{J} = \mathcal{J}_m + \mathcal{J}_{\phi} + \mathcal{J}_b
\ee
where $\mathcal{J}_m$, $\mathcal{J}_{\phi}$ and $\mathcal{J}_b$ correspond to the measurement, equations and boundary losses, respectively, 
\be
    \mathcal{J}_m =\frac{1}{2}\left\Vert\mathcal{M}\left(\boldsymbol{u}\right) - \boldsymbol{m}\right \Vert^2, 
\ee
\be
    \mathcal{J}_{\phi} = \lambda_c\mathcal{J}_c+\lambda_u\mathcal{J}_u+\lambda_v\mathcal{J}_v+\lambda_w\mathcal{J}_w, 
\ee
\be
    \mathcal{J}_b = \left\Vert \mathscr{B}\left(\boldsymbol{u},p\right)\right\Vert^2. 
\ee
The equations, or physics, losses are further detailed below, and are defined as the $L^2$ residuals of the incompressible Navier-Stokes equations (\ref{eq:NS}),
\be
    \begin{split}
        \mathcal{J}_c &= \left\Vert\mathscr{N}_c\left(\boldsymbol{u},p\right) \right\Vert^2 = \left\Vert\nabla \cdot \boldsymbol{u}\right\Vert^2\\
        \mathcal{J}_u &= \left\Vert\mathscr{N}_u\left(\boldsymbol{u},p\right) \right\Vert^2 =\left\Vert\frac{\partial u }{\partial t} +  \boldsymbol{u} \cdot \nabla u +  \frac{\partial p }{\partial x} - \frac{1}{\mathrm{Re}} \nabla^2 u \right\Vert^2\\
        \mathcal{J}_v &= \left\Vert\mathscr{N}_v\left(\boldsymbol{u},p\right) \right\Vert^2 =\left\Vert\frac{\partial v }{\partial t} +  \boldsymbol{u} \cdot \nabla v +  \frac{\partial p }{\partial y} - \frac{1}{\mathrm{Re}} \nabla^2 v \right\Vert^2\\
        \mathcal{J}_w &= \left\Vert\mathscr{N}_w\left(\boldsymbol{u},p\right) \right\Vert^2 =\left\Vert\frac{\partial w }{\partial t} +  \boldsymbol{u} \cdot \nabla w +  \frac{\partial p }{\partial z} - \frac{1}{\mathrm{Re}} \nabla^2 w \right\Vert^2. 
    \end{split}
\ee
In this approach, both the governing equations and boundary conditions are incorporated as $L^2$ regularizations into the loss function that attempts to reproduce the measurements.  Compared to the original data-assimilation problem (\ref{eq:NS optimization}), the PINN formulation (\ref{eq:WC optimization}) is an unconstrained optimization and can be solved directly using gradient type optimization methods. 

\begin{figure}\label{fig:schematics}
    \centering
    \begin{subfigure}{1.0\linewidth}
    \centering
    \includegraphics[width=1.0\textwidth]{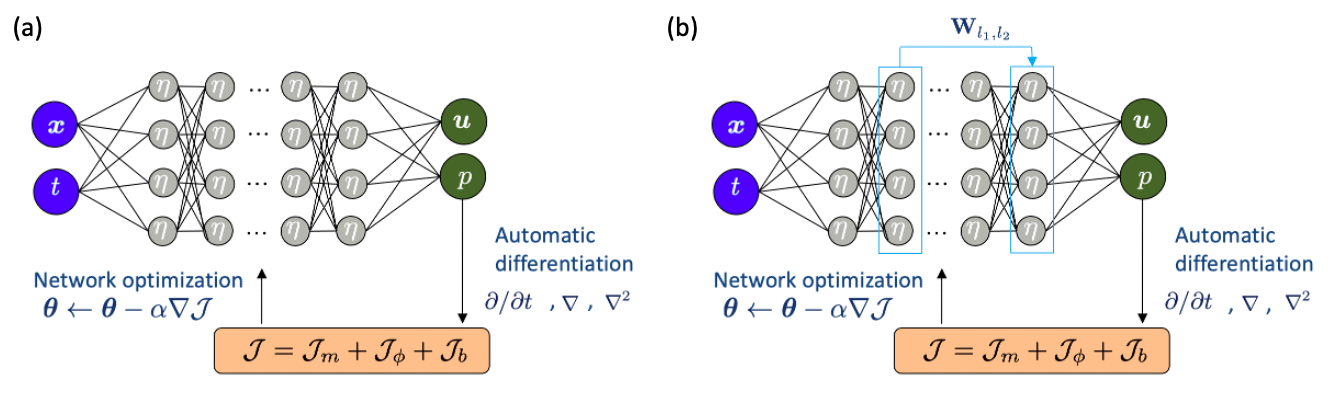}
    \end{subfigure}%
    \caption{Schematics of PINN structures. The networks represent the spatio-temporal solution, and is trained subject to the measurement, physics and boundary losses. (a) PINN with fully connected network.  (b) PINN with ResNet structure, where an additional connection is implemented between non-adjacent layers $l_1$ and $l_2$ with weights $\mathbf{W}_{l_1, l_2}$.}
    \label{fig:schematics pinn}
\end{figure}

The functions $(\boldsymbol{u}, p)$ are parameterized using a neural network that consists of an input layer, hidden layers with depth $D$, and an output layer (see figure \ref{fig:schematics pinn}).  
The input layer of this network is a column vector that contains the normalized spatial and temporal coordinates:
$$
	\mathbf{g}_0 = 
	(\tilde{x},
	\tilde{y},
	\tilde{z},
	\tilde{t})^{\top}
$$
where $\tilde{\bullet} = (\bullet - \overline{\bullet})/\left(\max{(\bullet)} - \min{(\bullet)}\right)$ is the centered and normalized coordinate. Each hidden layer contains $n_L$ neurons, denoted as $\mathbf{g}_l$, and adjacent layers are related by, 
\be
    \label{eq:NN_plain}
	\mathbf{g}_{l+1}(\mathbf{g}_l) = \eta(\mathbf{W}_l\mathbf{g}_l + \mathbf{b}_l), \quad 0 \leq l \leq D,
\ee
where the matrix $\mathbf{W}_l$ contains all the weights at the $l^{\textrm{th}}$ layer, vector $\mathbf{b}_l$ represents the biases, and $\eta(\bullet)$ is the activation function that acts element-wise on vectors. The vector containing $\mathbf{W}_l$ and $\mathbf{b}_l$ for all layers in the network will be denoted $\boldsymbol{\theta}$.  The output layer is defined as the normalized flow variables: 
$$
		\mathbf{g}_{D+1} = 
	\begin{pmatrix}
	\tilde{u}\\
	\tilde{v}\\
	\tilde{w}\\
	\tilde{p}
	\end{pmatrix}
	=
	\begin{pmatrix}
	(u-\mu_{u})/\sigma_u\\
	(v-\mu_{v})/\sigma_v\\
	(w-\mu_{w})/\sigma_w\\
	p/(\frac{1}{2}\rho u_{\tau}^2)
	\end{pmatrix}
$$
where the mean values $(\mu_{u}, \mu_{v}, \mu_{w})$ and standard deviations $(\sigma_{u}, \sigma_{v}, \sigma_{w})$ are calculated from the observations alone.  The field variables $\tilde{\boldsymbol{u}} = (\tilde{u}, \tilde{v}, \tilde{w})$ and $\tilde{p}$, and similarly $(\boldsymbol{u},p)$,  are all functions of $(x,y,z,t)$ and parameterized by $\boldsymbol{\theta}$. 
%   \del{; we therefore denote the PINNs fields as  $\hat{\boldsymbol{u}}_{\boldsymbol{\theta}}$ and $\hat{p}_{\boldsymbol{\theta}}$} 
In terms of this description, the minimization problem (\ref{eq:WC optimization}) can be reformulated as the search for the optimal $\boldsymbol{\theta}$: 
\be\label{eq:WC reparameterized}
    \min_{\boldsymbol{\theta}} \mathcal{J}\left(\boldsymbol{u},p\right).
\ee

We adopt the stochastic gradient descent method (SGD) to search for the optimal $\boldsymbol{\theta}$ that minimizes the loss function $\mathcal{J}$.  In SGD, the gradient of a mini-batched loss is calculated and used to update the network parameters: 
\be\label{eq:SGD}
	\boldsymbol{\theta}^{n+1} = \boldsymbol{\theta}^{n} - \alpha \nabla _{\boldsymbol{\theta}}\hat{\mathcal{J}}(\boldsymbol{\theta}^{n}),
\ee
where $\alpha$ is the learning rate, $\boldsymbol{\theta}^{n}$ is the $n^{\textrm{th}}$ iteration of the network parameters, $\hat{\mathcal{J}}$ is the mini-batch loss calculated from partial observation and random sample points for the equations and boundary conditions losses: 
\be
	\hat{\mathcal{J}} = \hat{\mathcal{J}}_m + \hat{\mathcal{J}}_{\phi} + \hat{\mathcal{J}}_b. 
\ee
The mini-batch losses $\hat{\mathcal{J}}_m$, $\hat{\mathcal{J}}_{\phi}$ and $\hat{\mathcal{J}}_b$ are defined as following
\be \label{eq:batch obs loss}
    \hat{\mathcal{J}}_m(\boldsymbol{\theta}) =\frac{1}{2}\left\Vert\mathbf{R}\left(\mathcal{M}\left(\boldsymbol{u}\right) - \boldsymbol{m}\right)\right \Vert^2,
\ee
\be\label{eq:batch physics loss}
    \hat{\mathcal{J}}_{\phi}(\boldsymbol{\theta}) = \lambda_c\hat{\mathcal{J}}_c+\lambda_u\hat{\mathcal{J}}_u+\lambda_v\hat{\mathcal{J}}_v+\lambda_w\hat{\mathcal{J}}_w,
\ee
\be\label{eq:batch boundary loss}
    \hat{\mathcal{J}}_b(\boldsymbol{\theta}) = \sum_{j=1}^{N_b}\left\vert \mathscr{B}\left[\boldsymbol{u},p\right](\boldsymbol{x}_{j}, t_{j})_{b}\right\vert^2.
\ee
In the above expressions, $\mathbf{R} \in \mathbb{R}^{N_{s}\times N_{m}}$ is a random matrix that has one element on each row equal to unity and the other elements equal to zero, which represents the random choice of observation data. The dimension $N_{m}$ is the number of observation data, and $N_{s}\ll N_{m}$ is the size of mini-batch data set. The batch boundary conditions loss is formed by sampling the boundary conditions mismatch $\mathscr{B}_k\left[\boldsymbol{u},p\right]$ on a set of random points $(\boldsymbol{x}_{j}, t_{j})_{b}$ that are generated on $\partial\Omega$. The batched physics loss $\hat{\mathcal{J}}_{\phi}$ is calculated as a linear combination of batched equations losses. For example, the batched continuity loss is defined as,
	\be \label{eq:batch physics loss terms}
    \hat{\mathcal{J}}_c(\boldsymbol{\theta}) = \sum_{j=1}^{N_p}\left\vert \mathscr{N}_c\left[\boldsymbol{u},p\right](\boldsymbol{x}_{j}, t_{j})_{\phi}\right\vert^2
\ee
where the points $\{(\boldsymbol{x}_{j}, t_{j})\}_{\phi}$ are generated randomly in the flow domain $\Omega$ using a uniform distribution. The batched loss in the momentum equations, $\hat{\mathcal{J}}_u, \hat{\mathcal{J}}_v$ and $\hat{\mathcal{J}}_w$, can be similarly defined on the randomly sampled points.
The sequence $\boldsymbol{\theta}^{n}$ generated from equation (\ref{eq:SGD}) converges in distribution to an optimal that minimizes the total loss (\ref{eq:WC reparameterized}). The SGD iteration, or training process, terminates when the loss stagnates, specifically when the relative decrease of loss is less than 10\% within $10^{4}$ epochs.  The reconstructed velocity and pressure fields at any spatial position and time can be obtained by providing the coordinates $(\boldsymbol{x},t)$ as input to the network and evaluating the output $(\boldsymbol{u}, p)$.

% \begin{figure}\label{fig:schematics}
%     \centering
%     \begin{subfigure}{0.5\linewidth}
%     \centering
%     \includegraphics[width=0.9\textwidth]{schematics/PINN_schematics1.png}
%     \caption{}
%     \end{subfigure}%
%     \begin{subfigure}{0.5\linewidth}
%     \centering
%     \includegraphics[width=0.9\textwidth]{schematics/PINN_schematics2.png}
%     \caption{}
%     \end{subfigure}
%     \caption{Schematics of PINN structures. The networks represent the spatio-temporal solution, and is trained subject to the measurement, physics and boundary losses. (a) PINN with fully connected network.  (b) PINN with ResNet structure, where an additional connection is implemented between non-adjacent layers $l_1$ and $l_2$ with weights $\mathbf{W}_{l_1, l_2}$.}
%     \label{fig:schematics pinn}
% \end{figure}

The relative weighting among different terms in the loss functions $\mathcal{J}$, and similarly $\hat{\mathcal{J}}$, is important since improper balance among these terms could lead to numerical stiffness in the training \citep{wang2020understanding}, lower test performance of the neural network, or an important entry (e.g.~one of the physical equations) not being satisfied with sufficient accuracy. 
Thus a dynamic strategy is adopted to reach a balance between different terms of the loss function and avoid manual tuning of the parameters. For example, the weight on the continuity equation, $\lambda_{c}$, is calculated at each iteration as, 
\be\label{eq:dynamic weighting}
    \lambda_{c} = \frac{\vert \nabla_{\boldsymbol{\theta}}\hat{\mathcal{J}}_m\vert_2}{\vert \nabla_{\boldsymbol{\theta}}\hat{\mathcal{J}}_{c}\vert_2}. 
\ee
This expression re-scales the gradient $\nabla \hat{\mathcal{J}_{c}}$ to the same $L^2$ magnitude as $\nabla \hat{\mathcal{J}}_{m}$. The weights for the other equation losses are calculated similarly.  

A summary of the PINN procedure is provided in Algorithm \ref{alg:PINN}, and the architecture of the network is shown schematically in figure \ref{fig:schematics pinn}.
The network expressivity must be commensurate with the complexity of chaotic, turbulent channel flow.  Deep networks are more suitable than shallow, wide networks \citep{poole2016exponential}. However, the optimization of deep networks are notoriously difficult due to the irregular loss function landscape \citep{li2018visualizing}. 
The ResNet architectures, or skip connections, can improve the training efficiency and performance, because this type of neural network behaves similar to an ensemble of shallow networks \citep{veit2016residual}. 
Two ResNet architectures proposed in \citep{he2016deep} are considered here.
The first, which we term ResNet1, is formulated as:
\be\label{eq:skip1}
	\mathbf{g}_{l_2}(\mathbf{g}_{l_2-1},\mathbf{g}_{l_1}) = \eta(\mathbf{W}_{l_2-1}\mathbf{g}_{l_2-1} + \mathbf{b}_l) + \mathbf{g}_{l_1},
\ee
where $l_1 < l_2-1$. 
Compared with the plain network (\ref{eq:NN_plain}), an extra term $\mathbf{g}_{l_1}$ is included to directly connect a layer $l_1$ to $l_2$.
The second architectures, ResNet2, modifies equation (\ref{eq:skip1}) by introducing a weight matrix in the  $\mathbf{g}_{l_1}$ term,
\be\label{eq:skip2}
	\mathbf{g}_{l_2}(\mathbf{g}_{l_2-1},\mathbf{g}_{l_1}) = \eta(\mathbf{W}_{l_2-1}\mathbf{g}_{l_2-1} + \mathbf{b}_l) + \mathbf{W}_{l_1,l_2}\mathbf{g}_{l_1},
\ee
where $\mathbf{W}_{l_1,l_2}$ is optimized during the SGD training.

\begin{algorithm}[h]
		\SetAlgoLined
		\textbf{Step 1}: Feed-forward evaluation\;
		\Indp
		\textbullet~Start with estimate of network parameters $\boldsymbol{\theta}^{n}$. For $n=0$, generate the weights $\mathbf{W}_{l}$ with standard normal distribution and biases $\mathbf{b}_l$ with zero values\;
		\textbullet~Generate a random copy of $\mathbf{R}$ in equation (\ref{eq:batch obs loss}). Evaluate the network at observation points sampled by $\mathbf{R}$. 
		Calculate batched observation loss $\hat{\mathcal{J}}_m$ in (\ref{eq:batch obs loss})\;
		\textbullet~Randomly sample the physics points $\{(\boldsymbol{x}_{j}, t_{j})_{\phi}\}$ in equation (\ref{eq:batch physics loss terms}). 
		Evaluate network at these points, and calculate the batched physics loss $\hat{\mathcal{J}}_{\phi}$ in (\ref{eq:batch physics loss})\;
		\textbullet~Randomly sample the boundary points $\{(\boldsymbol{x}_{j}, t_{j})_{b}\}$ in equation (\ref{eq:batch boundary loss}). Evaluate network at boundary points. Calculate the batched boundary loss $\hat{\mathcal{J}}_b$ in (\ref{eq:batch boundary loss})\;
		\Indm
		\textbf{Step 2}: Back propagation\;
		\Indp
		\textbullet~Using automatic differentiation, calculate the gradients of batched losses $\nabla\hat{\mathcal{J}}_m$,  $\nabla\hat{\mathcal{J}}_b$, $\nabla\hat{\mathcal{J}}_c$, $\nabla\hat{\mathcal{J}}_u$, $\nabla\hat{\mathcal{J}}_v$, $\nabla\hat{\mathcal{J}}_w$ \;
		\textbullet~Calculate the weights of batched equation losses $\lambda_{c}$, $\lambda_{u}$, $\lambda_{v}$, $\lambda_{w}$\;
		\textbullet~Calculate the total batched gradient $\nabla\hat{\mathcal{J}}$\;
		\textbullet~Update the network parameters $\boldsymbol{\theta}^{n}$ using equation (\ref{eq:SGD})\; 
		\textbullet~Repeat Steps 1-2 till the loss function reaches convergence criterion\;
		\Indm
		\textbf{Step 3}: Evaluate the flow trajectory\;
		\Indp
		\textbullet~Generate the spatio-temporal coordinates of interest in the flow field.\;
		\textbullet~Evaluate the network at these input points to obtain the velocity and pressure.\ 
		\caption{PINN state estimation.}
		\label{alg:PINN}
\end{algorithm}

In the above description, the boundary conditions and physics constraints were incorporated in the loss function. Some of the original constraints in (\ref{eq:NS optimization}) can, however, be embedded into the network structure in order to ensure that they are enforced exactly, to machine precision. For example, to embed the periodic boundary condition into the network structure, we can adopt an input-feature expansion \citep{yazdani2020systems,du2021ednn} where the first layer of the network is constructed as,  
\be\label{eq:periodic BC}
\mathbf{g}_0 = 
	\left(
	\sin\left(\frac{2\pi}{L_x}x\right), 
	\cos\left(\frac{2\pi}{L_x}x\right), 
	y, 
	\sin\left(\frac{2\pi}{L_z}z\right), 
	\cos\left(\frac{2\pi}{L_z}z\right), 
	t 
	\right)^{\top}. 
\ee
By this construction, the network function is automatically periodic in the horizontal $x$ and $z$ directions.  Another constraint that could be exactly embedded into the network is the continuity equation.  Instead of predicting the velocity and pressure, the network output is set to be a vector potential and pressure: 
\be\label{eq: vector potential}
	\mathbf{g}_{D+1} = \left(
	\Psi_x,
	\Psi_y,
	\Psi_z,
	\tilde{p} \right), 
\ee
where $\boldsymbol{\Psi}=(\Psi_x, \Psi_y, \Psi_z)$ is the vector potential of the incompressible flow field. The velocity field $\boldsymbol{u}=\nabla\times\boldsymbol{\Psi}$ is then evaluated using automatic differentiation. The resulting velocity field is automatically divergence free up to machine error.   

The majority of the results presented herein (\S\ref{sec:results_gen}-\S\ref{sec:results_basis}) are obtained using the PINN formulation where the constraints are weakly enforced through the loss function.  Results from PINNs that adopt embedded constraints will be reported in \S\ref{sec:results_constraints}.

%================================
%       RESULTS
%================================
\section{Results}
\label{sec:results} 

Using 4DVar and PINNs, we will examine the accuracy of estimation of turbulent channel flow from under-resolved observation data. The computational parameters for 4DVar are listed in tables \ref{table:adjoint} where the grid resolution is reported and matches that of the reference simulation.
The PINN architectures are summarized in table \ref{table:NN}, including the depth $D$ of the network and number of neurons per hidden layer $n_L$.  
% The abbreviation ``ResNet'' refers to skip-connection neural networks (figure \ref{fig:schematics pinn}b), 
%   \tz{It is important to remark that the dimensions of the network parameters to be optimized, $\mathrm{dim}(\boldsymbol{\theta})$, are an order of magnitude smaller than the dimension of the initial state, $\mathrm{dim}(\boldsymbol{u}_0)$, which is the object of the 4DVar optimization.  The efficiency of the PINN in representing the flow state is even more notable if we consider that the trained network represents the entire flow evolution during the observation horizon, not solely the initial condition.}
% \yd{All the state estimation cases focus on $\mathrm{Re}_{\tau} = 180$, except the F-AJ-392 and F-NN-S$_{\mathbf{W}}$-392 cases that investigate the performance of 4DVar and PINN at a higher Reynolds number $\mathrm{Re}_{\tau}=392$.}
% The discussion starts with a benchmark comparison between the state estimation using 4DVar and PINNs (\S\ref{sec:results_gen}), where the focus is placed on the convergence history and accuracy. 
%\yd{The robustness for the comparison of reconstruction accuracy against Reynolds number is briefly discussed at the end of \S\ref{sec:results_gen}.}
Benchmark results are reported in \S\ref{sec:benchmark} using fine observation data.  We proceed to study the influence of observation resolution on the estimation accuracy of both methods in \S\ref{sec:results_resolution}.   
In \S\ref{sec:results_basis}, we assess the impact of the network structure of PINNs, which is mathematically equivalent to adopting different functional basis to represent the flow fields.
The benefits of enforcing embedded constraints in PINNs are addressed in \S\ref{sec:results_constraints}. 
The results in this section are at $Re_\tau = 180$; the influence of a higher Reynolds number ($Re_\tau = 392$) on the estimation accuracy by 4DVar and PINN is examined in Appendix \ref{sec:appendix B}.

\begin{table}[b]
	\centering
		\begin{tabular}{l c c c c c c}
		    \hline
			 Case  & $\Delta x^+$   & $\Delta y^+$  & $\Delta z^+$ & $\Delta t^+$  & dim($\boldsymbol{u}_0$) & section \\
			\hline 
			 %R100 & 8 & 8 & 8 & 8  & 39.2 & [2.4, 9.6] & 19.6 & 2.2 &  49.26 \\
		\rowcolor{blue!20} 	 F-AJ & 4.9 & [0.4, 1.4] & 2.45 & 0.07 & $4.8 \times 10^6$ & \S3.1 \\
	     	\hline 
             C-AJ & 4.9 & [0.4, 1.4] & 2.45 & 0.07 & $4.8 \times 10^6$ & \S3.2 \\
			\hline 
			%F-AJ-392 & \yd{392} & 4.9 & [0.3, 1.3] & 2.45 & 0.07 & $1.14 \times 10^7$ & \S \ref{sec:appendix B} \\
			%\hline 
	\end{tabular}
	\caption{Parameters of 4Dvar state estimation. In the case designation, the first letter refers to (`F') fine or (`C') coarse measurements resolution (see table \ref{table:obs}). The letters `AJ' refer to 4DVar. The parameters $\Delta x^+$, $\Delta y^+$, $\Delta z^+$ and $\Delta t^+$ are the grid sizes and time step in viscous units.}
	\label{table:adjoint}
\end{table}

% \begin{table}
% 	\centering
% 		\begin{tabular}{l c c c c c c c c c }
% 		    \hline
% 	 	     Case & \yd{$\mathrm{Re}_{\tau}$} & $D$ & $n_{L}$ & structure & $\mathrm{dim}(\boldsymbol{\theta})$& section  \\
% 			\hline 
% \rowcolor{blue!20} 	 F-NN-S$_{\mathbf{W}}$ & \yd{180} & 10 & 200 & ResNet\del{1}\yd{2} & $4.04\times10^{5}$ & \S 3.1, 3.2, 3.3  \\ 
% \rowcolor{blue!20} 	C-NN-S$_{\mathbf{W}}$ & \yd{180} & 10  & 200 & ResNet\del{1}\yd{2} & $4.04\times10^{5}$ & \S3.2 \\

% 			\hline 
% 	    	 F-NN-V & \yd{180} & ~4  & 350 & Plain & $3.71\times10^{5}$ & \S3.3  \\
% 			 F-NN-W & \yd{180} & ~4  & 500 & Plain & $7.56\times10^{5}$ & \S3.3  \\
% 			 F-NN-S$_{\mathbf{I}}$ & \yd{180} & 10 & 200 & ResNet\del{2}\yd{1} & $3.63\times10^{5}$ & \S3.3  \\
% 			 F-NN-S$_{\mathbf{W}}$-EC & \yd{180} & 10 & 200 & ResNet\del{1}\yd{2} & $4.04\times10^{5}$ & \S 3.4  \\	      %F-NN-S$_{\mathbf{W}}$-392 & \yd{392} & 15 & 200 & ResNet1 & $6.85\times10^{5}$ & \S 3.1  \\
% 			\hline 
% 	\end{tabular}
% 	\caption{Parameters of PINNs. In the case designation, the first letter refers to (`F') fine or (`C') coarse measurements resolution (see table \ref{table:obs}). The letters `NN' indicate use of PINNs. The last entry is the network structure: (`V') vanilla, (`W') wide, (`S$_{\mathbf{W}}$', `S$_{\mathbf{I}}$') are two types of skip-connection networks. Parameter $D$ is the depth of hidden layers; $n_{L}$ is number of neurons per layer.}
% 	\label{table:NN}
% \end{table}

\begin{table}
	\centering
		\begin{tabular}{l c c c c c c c c }
		    \hline
	 	     Case & $D$ & $n_{L}$ & structure & $\mathrm{dim}(\boldsymbol{\theta})$& section  \\
			\hline 
\rowcolor{blue!20} 	 F-NN-S$_{\mathbf{W}}$ & 10 & 200 & ResNet2 & $4.04\times10^{5}$ & \S 3.1, 3.2, 3.3  \\ 
\rowcolor{blue!20} 	C-NN-S$_{\mathbf{W}}$ & 10  & 200 & ResNet2 & $4.04\times10^{5}$ & \S3.2 \\

			\hline 
	    	 F-NN-V  & ~4  & 350 & Plain & $3.71\times10^{5}$ & \S3.3  \\
			 F-NN-W  & ~4  & 500 & Plain & $7.56\times10^{5}$ & \S3.3  \\
			 F-NN-S$_{\mathbf{I}}$  & 10 & 200 & ResNet1 & $3.63\times10^{5}$ & \S3.3  \\
			 F-NN-S$_{\mathbf{W}}$-EC  & 10 & 200 & ResNet2 & $4.04\times10^{5}$ & \S 3.4  \\	      %F-NN-S$_{\mathbf{W}}$-392 & \yd{392} & 15 & 200 & ResNet1 & $6.85\times10^{5}$ & \S 3.1  \\
			\hline 
	\end{tabular}
	\caption{Parameters of PINNs. In the case designation, the first letter refers to (`F') fine or (`C') coarse measurements resolution (see table \ref{table:obs}). The letters `NN' indicate use of PINNs. The last entry is the network structure: (`V') vanilla, (`W') wide, (`S$_{\mathbf{W}}$', `S$_{\mathbf{I}}$') are two types of skip-connection networks. Parameter $D$ is the depth of hidden layers; $n_{L}$ is number of neurons per layer.}
	\label{table:NN}
\end{table}

%-----------------------------
%       General results
%-----------------------------
\subsection{Benchmark comparison of 4DVar and PINNs}
\label{sec:benchmark}
% Most accurate estimation of turbulence: benchmark comparison of 4DVar and PINNs
% Optimal estimation of turbulence: benchmark comparison of 4DVar and PINNs
% Best possible estimation of turbulence using 4DVar and PINNs
\label{sec:results_gen}

The first comparison of 4DVar and PINN corresponds to cases F-AJ (table \ref{table:adjoint}) and F-NN-S$_{\mathbf{W}}$ (table \ref{table:NN}), the latter being a skip-connection network. 
These cases are selected because they provide a sense of the most accurate state estimation that was achieved in our setup, and hence will provide a backdrop for assessment of different parameters (e.g. coarser observations, different network architectures,...etc).  
%   In this specific test, we are adopting the fine observations, as denoted by the starting letter `F' in both case designations. 
In this benchmark test, we are adopting fine-resolution observations, as denoted by the starting letter `F' in both case designations.

%   \hl{(Major comment 1.1) }\yd{The numerical computation in cases F-AJ and F-NN-S$_{\mathbf{W}}$ are performed on different platforms with different computational costs. Specifically, the PINN training in case F-NN-$S_{\mathbf{W}}$ is performed on a NVIDIA A100 GPU with 6912 CUDA cores. The whole training process includes $3\times 10^{4}$ epochs, which in total takes 25.1 hours. Each epoch takes approximately $3.013$ seconds. The 4DVar calculation of case F-AJ is performed on Intel Cascade Lake 6248R CPU code using 24 CPU cores. The full 4DVar optimization procedure consists of 100 forward-adjoint iterations which takes 6.4 hours in total. Each forward-adjoint iteration takes 231 seconds. Since the computational efficiency of single GPU and multiple CPUs cannot be compared directly, the computational cost of both methods are documented only for the completeness of the work rather than a direct comparison of efficiency for both methods. }

% \del{The adjoint case uses a fine mesh to estimate the field, and the PINNs uses a skip-connection network (the effect of the network architecture will be revisited in detail in \S\ref{sec:results_basis}).}
%   \mw{The skip-connection network is adopted for PINNs (case F-NN-S$_{\mathbf{W}}$ in table \ref{table:NN}), and the effect of the network architecture will be revisited in detail in \S\ref{sec:results_basis}.}

\begin{figure}
    \centering
    \includegraphics[width=1.0\textwidth]{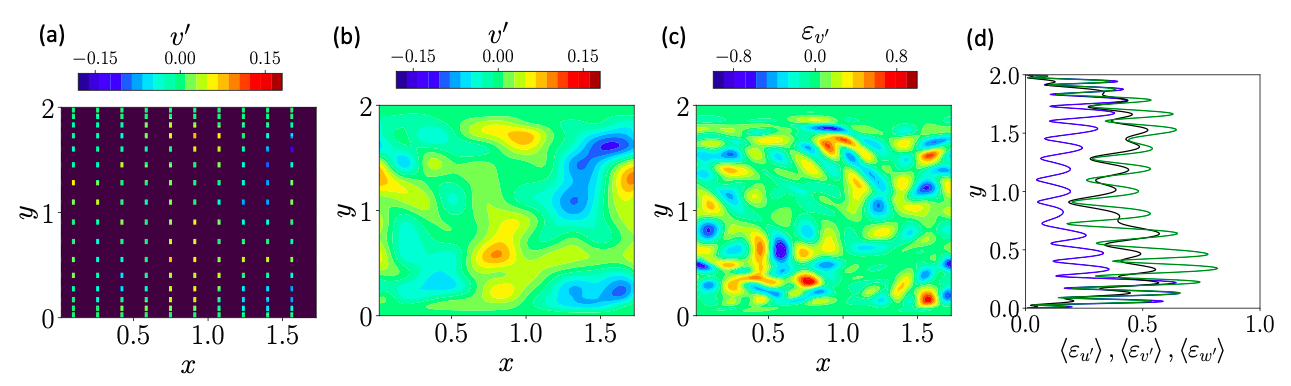}
    \caption{The velocity component $v'$ of initial guess $\boldsymbol{u}_0$ of 4DVar. (a): Sparse observations of $v'$. (b): The spatial interpolation of sparse observations from figure (a). (c): The error $\varepsilon_{v'}$ of the interpolated $v'$ against true initial condition $v'_r$. (d): The x-z averaged errors $\left<\varepsilon_{u'}\right>$ (\fullblue), $\left<\varepsilon_{v'}\right>$(\fullblack) and  $\left<\varepsilon_{w'}\right>$(\fullgreen) of interpolated field at initial time. }
    \label{fig:init guess}
\end{figure}

\begin{figure}
    \centering
    \begin{subfigure}{1.0\linewidth}
    \centering
    \includegraphics[width=1.0\textwidth]{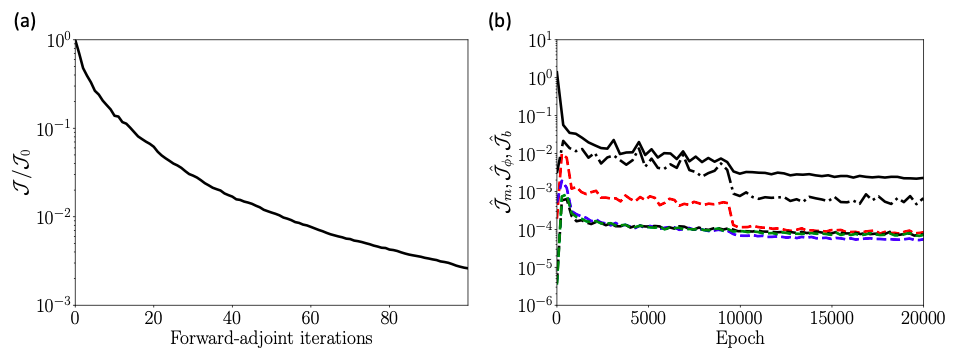}
    \label{fig:loss decay adjoint}
    \end{subfigure}%
    \caption{Convergence history of 4DVar and PINN.  (a): Total loss function normalized by its initial value versus number of forward-adjoint iterations. (b): Losses of PINNs versus number of training epochs, \fullblack: $\hat{\mathcal{J}}_{m}$,\chainblack: $\hat{\mathcal{J}}_{b}$,  \dashedred{}: $\hat{\mathcal{J}}_{c}$, \dashedblue{}: $\hat{\mathcal{J}}_{x}$, \dashedblack{}: $\hat{\mathcal{J}}_{y}$, \dashedgreen{}: $\hat{\mathcal{J}}_{z}$.}
    \label{fig:loss decay}
\end{figure}

A visualization of the sub-sampled vertical-velocity observations is shown in figure \ref{fig:init guess}(a).  Spatial interpolation of the observations at the initial time (figure \ref{fig:init guess}(b)) provides the first guess for the 4DVar optimization.  The spatial distribution of the errors in the interpolated field is shown in figure \ref{fig:init guess}(c), where $\varepsilon_{v'}=(v'-v'_r)/\langle v^{\prime}_r\rangle_{\Omega}$ and $\langle v^{\prime}_r\rangle_{\Omega}$ is the volume-averaged root-mean-square fluctuation at the initial time.
%  This estimation of initial condition captures the large scale structures of the flow field, yet the fine structures are lost. 
Large values of $\varepsilon_{v'}$ concentrate in small regions compared to the size dominant structures of the flow. 
The root-mean-squared errors averaged on the horizontal plane,  $\left<\varepsilon_{\phi}\right>_{xz}$,
% $\sqrt{\int \varepsilon_{\phi}^2\mathrm{d}x\mathrm{d}z/L_x L_z}$ 
are reported in figure \ref{fig:init guess}(d), and are on the order of $\{30\%, 40\%, 50\%\}$ for $\{u', v', w'\}$.  The 4DVar method starts from this initial guess with loss $\mathcal{J}_{m,0}$, and improves the estimation of the initial condition using the forward-adjoint iteration procedure.
The convergence history of the normalized loss $\mathcal{J}_m/\mathcal{J}_{m,0}$ is reported in figure \ref{fig:loss decay}(a), and decays monotonically. 
After 100 forward-adjoint loops, the loss function drops more than two orders of magnitudes.

The loss functions for PINN are shown in figure \ref{fig:loss decay}(b).  
The convergence history reflects the training process of the network. 
Each epoch on the horizontal axis represents a gradient-descent iteration over all data points. 
The initial learning rate $\alpha = 10^{-3}$ was reduced to $10^{-4}$ at the $10^{4}$th epoch, and further decreased to $10^{-5}$ at the $2\times 10^4$th epoch in order to ensure that $\alpha$ is sufficiently small for convergence of the training process.
Due to the randomness introduced in the stochastic gradient descent, the loss functions are oscillatory with respect to training epochs.  Overall the observation loss $\hat{\mathcal{J}}_m$ decreases monotonically, which indicates that the network predictions at the observation locations converge to the data through the whole training process.  In contrast, the boundary loss $\hat{\mathcal{J}}_b$ and equations losses $\left(\hat{\mathcal{J}}_c, \hat{\mathcal{J}}_u, \hat{\mathcal{J}}_v, \hat{\mathcal{J}}_w\right)$ first increase to a large value then slightly decrease through the training epochs.  
This trend is consistent with the initialization of the network where the weights $\mathbf{W}_l$ are set to be independent random numbers and the biases $\mathbf{b}_l$ are zero. As a result, based on the law of large numbers, the predictions $(\boldsymbol{u},p)$ and their gradients are nearly zero relative to the reference flow scales; therefore the physics residual $\mathcal{J}_{\phi}$ is small relative to the leading order terms in the equations for true flow. The training process modifies the outputs to match the data, and the equations are no longer satisfied, which leads to the increase in the loss.  Further training enforces the equations and boundary conditions, which can be regarded as regularizations terms, in the later stages of training. At the end of the training procedure, all the losses have reduced and stagnate at a low level, which indicates that the network approximates the observations and satisfies, in the weak sense, the governing equations and boundary conditions.

The different convergence history of 4DVar and PINN is not merely due to the difference in the optimization procedure, but very importantly due to a difference in the loss-functions landscapes.
For 4DVar, two key challenges are the exponential amplifications of (i) errors in the initial state during the forward evolution and (ii) errors in the measurements during adjoint evolution.  These challenges are mitigated since the observation horizon is within one Lyapunov timescale, and the gradient-based L-BFGS algorithm is effective at monotonically reducing the loss during successive iterations (c.f. figure \ref{fig:loss decay}(a)).  When a longer observation horizon is of interest, multiple assimilation windows can be adopted, each shorter than the Lyapunov timescale.  Most importantly, in 4DVar all the considered trajectories in state space are solutions to the Navier-Stokes equations.  This property should be contrasted to the PINNs approach, where the physics are not a hard constraints but rather part of the loss.  It is therefore expected that the training of the deep PINN, which is required to express both the data and solution losses, contends with a much more tortuous landscape.  For this reason, SGD is adopted and the convergence history is non-monotonic (figure \ref{fig:loss decay}(b)).

\begin{figure}
    \centering
    \begin{subfigure}{0.92\linewidth}
    \centering
    \includegraphics[width=1.0\textwidth]{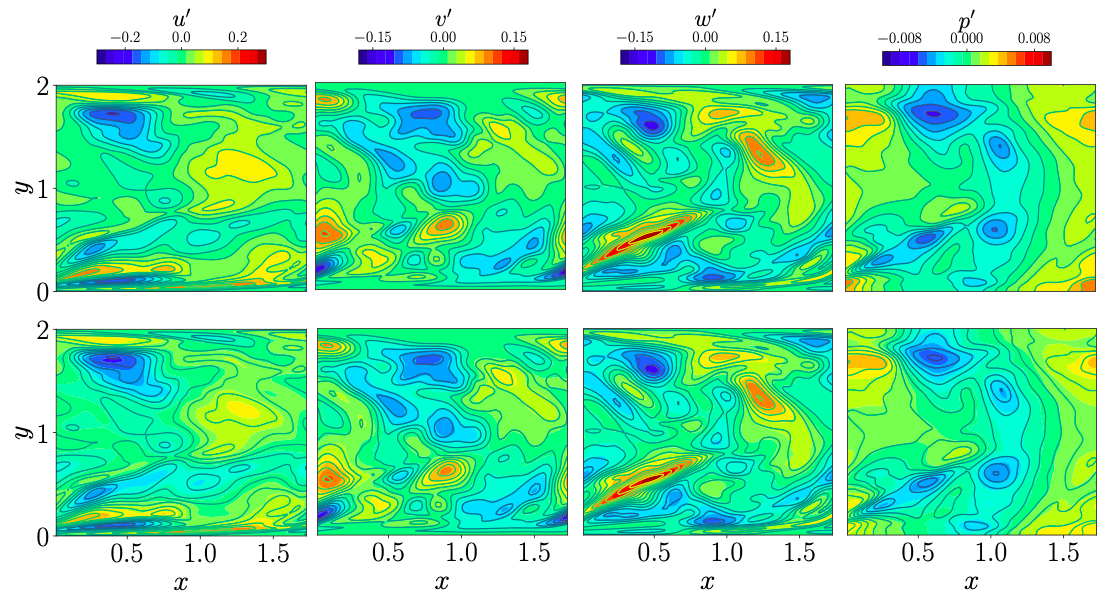}
    \end{subfigure}
    \caption{Observations and reconstructed fields at $t = T$, $z = L_z/2$. 
    % \del{Top row: sub-sampled velocity observations. Middle row:} 
    Top row: color contours show 4DVar estimated fields with line contours for the true solutions. Bottom row: color contours show PINN estimated fields with line contours for the true solutions. Columns from left to right are $u'$, $v'$, $w'$ and $p'$. }
    \label{fig:2d contour}
\end{figure}

A visual representation of the accuracy of the flow estimation is helpful, and is provided in figure \ref{fig:2d contour} at the end of the assimilation window $t=T$ when the adjoint reconstruction is the most accurate\citep{wang2021state}.  
The contours show the fluctuations of the three velocity components and pressure, $(u',v',w',p')$, all evaluated by subtracting the true mean from the estimated fields, for example 
\be
    u'(x,y,z)=u(x,y,z)-\overline{u}_r(y), \quad \, \, \overline{u}_r(y) = \frac{1}{TL_xL_z}\int_{0}^{T} \int_{0}^{L_x} \int_{0}^{L_z}u_r(x,y,z,t)\mathrm{d}x\mathrm{d}z\mathrm{d}t. 
    \label{eq:fluctuation}
\ee
%   \del{While in general $(v,w)$ should not have a mean component and therefore $(v,w)=(v',w')$, we avoid making this assumption and verify its accuracy in the PINN prediction.  }
%   We should remark that the mean flow of the estimated state within the assimilation window does not match the true mean, 
%e.g. $\overline{v}_r = 0$ but $\max_y\overline{v}(y)=1.39\times 10^{-4}$ 
% \max_y\overline{w}(y) = 2.40 \times 10^{-2}
%for PINN prediction, 
This definition lumps any deviations between the estimated and true means in the fluctuations reported in figure \ref{fig:2d contour}.
Such deviations are generally small; for example, instead of $\overline{v}_r = 0$, the PINN predicts $\overline{v} = 1.39 \times 10^{-4}$ which is two orders of magnitudes smaller than the maximum r.m.s.~fluctuations $\max_y v'_{rms} = 6.2 \times 10^{-2}$.
The top row of figure \ref{fig:2d contour} shows the observation data. The second row compares the 4DVar estimated fields (color contours) to the hidden truth (lines). The excellent agreement between the color and line contours indicates accurate estimation. The third row has the same qualitative comparison to the truth when the fields are estimated by the PINN.  While the large features are reproduced, some mismatch can be identified in small regions.  
This initial comparison suggests that the adjoint approach reconstructs the velocity and pressure fields with higher accuracy than PINNs\textemdash a conclusion that we will quantify.

Before evaluating the errors in the reconstruction, it is important to highlight that both 4DVar and PINNs were able to accurately reconstruct the pressure (see figure \ref{fig:2d contour}), which was not part of the measurements.  This prediction is therefore only possible by virtue of enforcing the governing equations (\ref{eq:NS}), strongly in the case of 4DVar and weakly, with sufficient accuracy, in the case of PINNs.

\begin{figure}[t]
    \centering
    \begin{subfigure}{0.92\linewidth}
    \centering
    \includegraphics[width=1.0\textwidth]{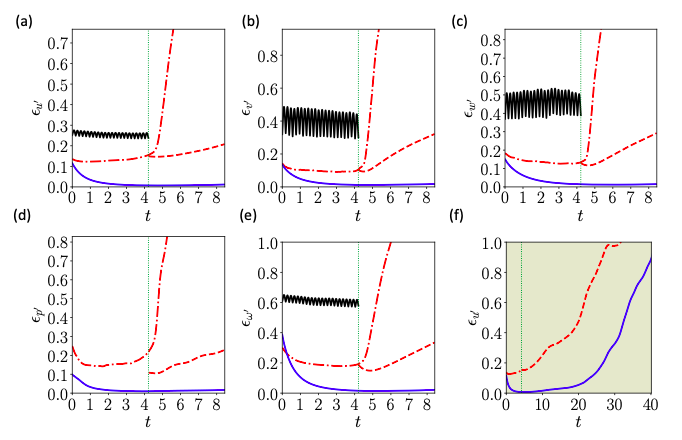}
    \end{subfigure}
    \caption{Estimation errors when observations are separated by one Taylor microscale. (a-e) $L^2$ error of estimated perturbation velocities, pressure and vorticity vector. \fullblack{}: Spatio-temporal interpolation,  \full{}: adjoint estimation case F-AJ, \chainred{}: PINNs estimation case F-NN-S$_{\mathbf{W}}$. \dashedred{}: DNS prediction using PINN solution at $t=T$ as initial condition.  
    (f) $L^2$ error of perturbation velocities $u'$ in an extended time horizon. Both forecasts are performed using DNS, starting at $t=T$ from (\full{}) the 4DVar estimate and (\chainred{}) the PINNs estimate.
    }
    \label{fig:1Lambda_error}
\end{figure}

In order to quantify the accuracy of the reconstructed states, we compute the normalized root-mean-squared error, 
\be\label{eq: rms error}
    \epsilon_{\psi}(t) = \sqrt{\frac{\int_{\Omega}\left\vert\psi(\boldsymbol{x},t)-\psi_r(\boldsymbol{x},t)\right\vert^2\mathrm{d}\boldsymbol{x}}{\int_{\Omega}\left\vert\psi_r(\boldsymbol{x},t)\right\vert^2\mathrm{d}\boldsymbol{x}}},
\ee
where averaging is performed over the channel volume,
%   \del{where $\psi$ is the reconstructed field, and $\psi_r$ is the reference solution, or hidden truth.}
$\psi$ denotes the estimated quantity, and $\psi_r$ is its reference, or true, value. 
The symbol $\vert \psi \vert$ denotes the absolute value when $\psi$ is a scalar and $L^2$ norm when $\psi$ is a vector.
The errors in the estimated velocity, pressure and vorticity disturbances are reported in figure \ref{fig:1Lambda_error}.  Note that the PINN estimate of vorticity, $\boldsymbol{\omega} = \boldsymbol{\nabla} \times \boldsymbol{u}$, was computed using automatic differentiation of the output velocity field with respect to the input spatial coordinates. 
The errors from the adjoint  (F-AJ) and PINN (F-NN-S$_{\mathbf{W}}$) are compared to spatio-temporal interpolation of the observations in the figure.  For the velocity (figure \ref{fig:1Lambda_error}a-c), interpolation errors are on the order of $40\%$ of the r.m.s.~fluctuations.  While the PINN reduces the errors to approximately $15\%$, the most accurate prediction is from 4DVar estimation (less than 5\% errors).
Although the observations do not include any pressure or velocity-gradient data, the PINN is able to estimate the pressure and vorticity with errors on the order of 20\%, again appreciably exceeding the errors from 4DVar.  A comparison of estimation errors from 4DVar and PINNs at higher Reynolds number, $\mathrm{Re}_{\tau}=392$, is provided in Appendix \ref{sec:appendix B}, with qualitatively similar results.
%   These small errors are consistent with the qualitative agreement between true solutions and the estimations in figure \ref{fig:2d contour}.

Despite the relatively lower accuracy of PINN, it is important to remark that the dimensions of the network parameters to be optimized, $\mathrm{dim}(\boldsymbol{\theta})=4.04 \times 10^5$, are an order of magnitude smaller than the dimensions of the initial state, $\mathrm{dim}(\boldsymbol{u}_0)=4.7 \times 10^6$, which are the object of the 4DVar optimization (see tables \ref{table:adjoint} and \ref{table:NN}). The efficiency of the PINN in representing the flow state is even more notable if we consider that the trained network represents the entire flow evolution during the observation horizon, not solely the initial condition.
% As such, improved accuracy of the PINN can be achieved by increasing the network size, which would naturally increase the training cost.
It is therefore natural to consider increasing the network size in order to improve the accuracy of PINN prediction. However, our experience at lower Reynolds number indicates that this strategy is not necessarily effective. 
%   Further comments on this point are provided in  \S\ref{sec:results_resolution}.
Here we simply wish to emphasize the efficiency of PINN in representing the solution trajectory.

In addition to comparing the accuracy of the adjoint and PINNs, the time-dependence of the errors should also be contrasted.  The adjoint errors decay in forward time, $t\in[0,T]$, reaching their lowest level at the end of the time horizon.  This behavior is typical of 4DVar:  The optimization of the initial condition is focused on minimizing the errors at the final time because the adjoint system is forced by the difference between the true measurements and their estimate, and amplifies exponentially at the Lyapunov rate as it is marched back in time. As a result, the latest errors in reproducing the measurements have the largest impact on the adjoint field at $t=0$, and hence the gradient of the loss function (equation \ref{eq:grad}), which is used to update the initial condition.   
The PINNs predictions are agnostic to time, since all the measurements and equations are considered instantaneously rather than along a trajectory in state space.
%   In the observation time window $[0,T]$, reconstruction error of adjoint decays along time due to the positive Lyapunov exponents in the adjoint system. In contrast the reconstruction error of PINNs remain uniform through the time window $[0,T]$. 

Figure \ref{fig:1Lambda_error} also shows forecasts from the 4DVar and PINNs approaches for times that exceed the assimilation horizon, i.e.\,beyond the final observation. Specifically, during the interval $t \in [T,2T]$ the adjoint prediction is obtained by continuing to march the Navier-stokes equations, while the PINN predictions are obtained by evaluating the network during this window $[T,2T]$ without further training.  
The 4DVar yields accurate forecasts, which indicates that the predicted trajectory in state space shadows the true flow.  It should be noted, however, that small deviations which are invariably present in the estimation are expected to amplify exponentially in time, and eventually lead to departure from the truth when the simulations are carried out for longer times (see figure \ref{fig:1Lambda_error}(f) for a much longer forecast).
The key point in figures \ref{fig:1Lambda_error}(a-e) is, however, the behavior of the PINN errors which amplify dramatically immediately beyond the observation horizon, $t > T$.  The rapid deterioration in accuracy is because the training of the PINN was only performed during the observation horizon. As such, the network predictions are not constrained in any fashion to satisfy the governing equations beyond $t=T$.   
%
%---- We removed the interpolation curve-----
%   The error of prediction by marching the interpolated field first decreases in a short time window then increases indefinitely. The initial decrease of error from the prediction of interpolated solution is due to the fact that the Navier-Stokes forward marching smoothes out the components in the interpolated solution that does not satisfy the Navier-Stokes equations. The increase of error in the later stage comes from the sensitivity of solution trajectory with respect to errors in the initial condition. 

Improving the accuracy of the PINN forecast can be accomplished in a number of ways.  Firstly, we can expand the original time horizon of the PINN training. This approach would require progressively larger networks for longer forecasts, with sufficient complexity to express the dynamics of the turbulence over the horizon of interest; The associated training would also become progressively more challenging and computationally expensive.  Another approach is to train a separate PINN for the forecast time horizon, $t\in[T,2T]$. The total loss for this new PINN is comprised of the equations, boundary conditions and the estimated field at $t=T$ as data.  While this approach is feasible, it is computationally inefficient because the PINN solves an optimization problem over the entire spatio-temporal domain rather than evolve the spatial solution of the governing equations in time. 

The most expedient and accurate approach to forecast the flow beyond the PINN observation horizon is to perform direct numerical simulations, with the initial condition taken as the PINN-estimated field at $t=T$.  
The DNS can be performed using conventional numerical approaches or the recently invented evolutional deep neural network (EDNN,\citep{du2021ednn}) which can accurately solve partial differential equations.  While we adopt the former, we remark on EDNN for the benefit of the interested reader: EDNN represents the flow field in space, at a given time, e.g.\,the estimated field from PINN at $t=T$; The governing Navier-Stokes equations, re-expressed in terms of the network parameters, are then used to update the state of the network in time, and thus the updated EDNN parameters predict the evolution of the flow. 

We performed DNS starting from the PINN-estimated field at $t=T$, and 
the evolution of the errors are shown by the red dashed lines in figure \ref{fig:1Lambda_error}. The errors in the velocity and vorticity initially reduce slightly because the DNS algorithm projects the terminal PINN field onto a Navier-Stokes solution.  Most notable is the sharp, discontinuous drop of the pressure error at $t=T$, because the PINN solution does not satisfy the divergence-free condition in contrast to the DNS algorithm.  
% Specifically, since the velocity field predicted by PINN is not exactly divergence free, a source term $(\partial/\partial t-\nu\nabla^2)(\nabla \cdot \boldsymbol{u})$ appears in the Poisson equation governing the pressure, which introduces error into the reconstructed pressure. At the first time step of DNS prediction, this source term is removed by the projection step in the fractional-step method, leading to sudden decrease of error on pressure. 
At later times ($t > T$) the reconstructed velocity, vorticity and pressure progressively deviate from the true state.  
Figure \ref{fig:1Lambda_error}(f) compares the long-time forecasts ($t \gg T$) performed using DNS of (blue) the 4DVar and (dashed red) PINN terminal estimated states; The former sustains appreciably better accuracy prior to the inevitable exponential divergence from the true state.

\begin{figure}
    \centering
    \begin{subfigure}{1.0\linewidth}
    \centering
    \includegraphics[width=1.0\textwidth]{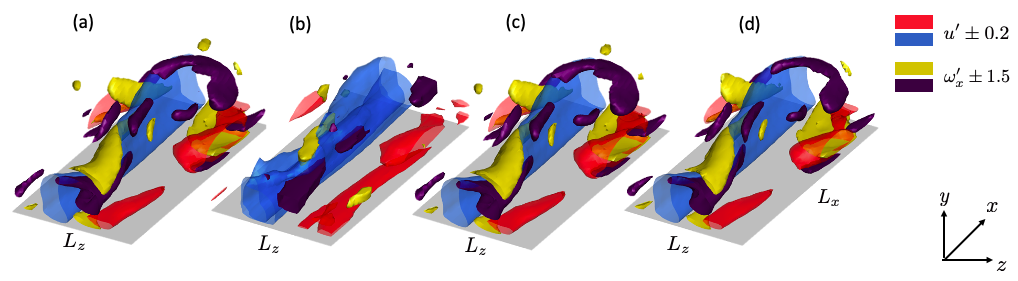}
    % \caption{Iso-contour of $\omega_x$ and $u'$ with one Taylor-microscale observation gap. (a) hidden truth (b) interpolation (c) adjoint estimation (d) PINNs}
    \end{subfigure}
    \caption{Iso-surfaces of $u'$ and $\omega'_x$ at $t=T$, when observations are separated by one Taylor microscale. (a) Hidden truth; (b) spatio-temporal interpolation; (c) 4DVar estimation case F-AJ; (d) PINN estimation case F-NN-S$_{\mathbf{W}}$.}
    \label{fig:3d contour 1Lambda}
\end{figure}

The estimated turbulent structures are shown in figure \ref{fig:3d contour 1Lambda}, specifically iso-surfaces of the streamwise velocity and vorticity perturbations.  The panels correspond to the true state, interpolated field, adjoint estimation and PINN reconstruction. 
The interpolated state is included to demonstrate that the observation data partially capture the velocity structures but miss the vorticity almost entirely.  Both 4DVar and PINN reproduce the velocity structures.   They can also discover, or estimate, the vorticity field, with qualitatively similar levels of fidelity although the 4DVar estimation is quantitatively more accurate (c.f.~errors in figure \ref{fig:1Lambda_error}).  
%   The vorticity of PINNs reconstruction $\omega = \nabla \times \boldsymbol{u}$ is calculated from automatic differentiations. Both the adjoint method and PINNs reconstruct the streak structure near the wall. The adjoint and PINN enforce governing equations (\ref{eq:NS}) in the whole domain, thus the velocity gradients are regularized by the governing equations. 

It is important to remark on the computational cost that underlies the performance of the algorithms for 4DVar (case F-AJ) and PINN (case F-NN-S$_{\mathbf{W}}$). 
The former was performed on Intel Cascade Lake 6248R CPU node using 24 CPU cores. Each forward-adjoint iteration required 231 seconds, and therefore the full 4DVar optimization procedure (100 forward-adjoint iterations) required 6.4 hours of wall-clock time.  
The PINN was executed on an NVIDIA A100 GPU with 6912 CUDA cores. Each training epoch required approximately $3.013$ seconds, and therefore the entire training process ($3\times 10^{4}$ epochs) required 25.1 wall-clock hours.  
Since the two algorithms were executed on different hardware architectures, their relative computational costs or performance is subjective. One may wish to compare the total wall-clock time, CPU core hours, energy consumption or hardware cost. While we leave this choice to the reader, here we  will simply remark that the better accuracy of the 4DVar algorithm was not due on an excessively high computational cost.

%   The higher accuracy of 4DVar compared to PINNs is due to several factors.
Several factors contribute to the higher accuracy of 4DVar compared to PINNs.
Firstly, in the adjoint approach the governing equations are enforced by marching an estimated initial condition forward in time, thus the evolution satisfies the Navier-Stokes equations to within the accuracy of the numerical discretization scheme.  The PINN, in contrast, enforces the governing equations approximately by minimizing an $L^2$ norm within the total loss function that comprises other contributions including the observations.  After training, the residual of the governing equations is finite and directly influences the fidelity of the estimated flow in particular away from the sparse observation points. 
Secondly, the observations that were used in this estimation exercise were generated using the same direct numerical simulation algorithm that was adopted in 4DVar.  In other words, the observations were generated by the same function space spanned by the 4DVar forward model, which brings a natural advantage to 4DVar in this comparison.  
% The realizations of differential operators in the governing equations for adjoint method and data acquisition model are based on the same piecewise polynomial basis on the same grid points. The residual of governing equations calculated from true solutions using the differential operators of forward model is exactly zero.  
To clarify this point, consider a scenario where the observations are fully resolved velocity and pressure fields of the true state, and we train the neural network by minimizing the measurement loss only. Such an estimated solution is  mathematically defined as,
\be
    (\boldsymbol{u}_{o}, p_{o}) = (\boldsymbol{u}_{\boldsymbol{\theta}_{o}}, p_{\boldsymbol{\theta}_{o}}), 
    \quad \, \, \boldsymbol{\theta}_{o} = \arg\min \left(\left\Vert \boldsymbol{u}_{\boldsymbol{\theta}}-\boldsymbol{u}_r\right\Vert+\left\Vert\boldsymbol{p}_{\boldsymbol{\theta}}-\boldsymbol{p}_r\right\Vert \right),
\ee
where $(\boldsymbol{u}_{\boldsymbol{\theta}_{o}}, p_{\boldsymbol{\theta}_{o}})$ explicitly specifies the dependence of the velocity and pressure on the network parameters $\boldsymbol{\theta}_{o}$. This optimal solution would not satisfy the governing equations calculated by automatic differentiation, since the differential operators calculated from finite volume and from automatic differentiation are different. The highest accuracy PINN could achieve depends on not only the representation capability of network, but also the difference between the realization of differential operators from finite volume method and automatic differentiation. 
The above factors justify the reduction in accuracy in the PINN estimates relative to 4DVar, and only the first concern regarding the weak enforcement of the governing equations should be of consideration in application to experimental measurements. 

% \tz{Despite the relatively lower accuracy of PINN, it is important to remark that the dimensions of the network parameters to be optimized, $\mathrm{dim}(\boldsymbol{\theta})=4.04 \times 10^5$, are an order of magnitude smaller than the dimensions of the initial state, $\mathrm{dim}(\boldsymbol{u}_0)=4.8 \times 10^6$, which are the object of the 4DVar optimization (see tables \ref{table:adjoint} and \ref{table:NN}). The efficiency of the PINN in representing the flow state is even more notable if we consider that the trained network represents the entire flow evolution during the observation horizon, not solely the initial condition. As such, improved accuracy of the PINN can be achieved by increasing the network size, which would naturally increase the training cost.}

%-----------------------------
%   Influence of resolution
%-----------------------------
\subsection{Influence of measurements resolution}
\label{sec:results_resolution}

\begin{figure}
    \centering
    \begin{subfigure}{0.9\linewidth}
    \centering
    \includegraphics[width=1.0\textwidth]{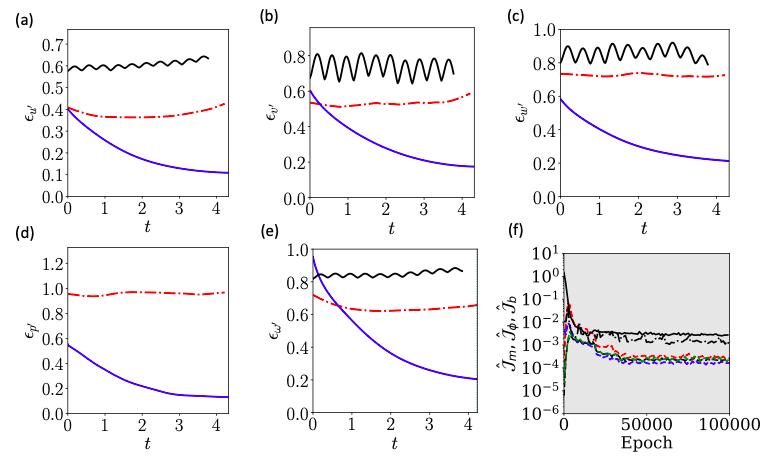}
    \end{subfigure}
    \caption{Estimation error and training history when observations are separated by two Taylor microscales. 
    (a-e) $L^2$ error of estimated perturbation velocities, pressure and vorticity vector. 
    \fullblack{}: Spatio-temporal interpolation,  \full{}: adjoint estimation case C-AJ, \chainred{}: PINNs estimation case C-NN-S$_{\mathbf{W}}$. 
    (f) Losses of PINN versus number of training epochs. \fullblack: $\hat{\mathcal{J}}_{m}$,\chainblack: $\hat{\mathcal{J}}_{b}$,  \dashedred{}: $\hat{\mathcal{J}}_{c}$, \dashedblue{}: $\hat{\mathcal{J}}_{x}$, \dashedblack{}: $\hat{\mathcal{J}}_{y}$, \dashedgreen{}: $\hat{\mathcal{J}}_{z}$.}
    \label{fig:2Lambda_error}
\end{figure}

The influence of the resolution of the observation on the accuracy of flow estimation is examined in this section.  In contrast to case `F' where the observations were spatially separated by one Taylor microscale (\S\ref{sec:results_gen}), we here consider nearly twice the spacing in case `C'. The estimation errors are reported in figure \ref{fig:2Lambda_error}, for both 4DVar and PINN, again compared to the spatio-temporal interpolation of the observations. 
In general, the errors have increased relative to the earlier predictions from the fine data (c.f.\,\ref{fig:1Lambda_error}).  The adjoint-variational approach yields the most accurate predictions at the end of the observation horizon: the errors in the velocity and pressure fields are all below $20\%$ of the r.m.s.~fluctuations.
In contrast, the PINNs predictions maintain a constant level of accuracy within the observation horizon, as anticipated.  For the velocities, these errors are higher than 4DVar and lower than spatio-temporal interpolation, although similar in order of magnitude to the latter for the $v'$ and $w'$ components. The accuracy of the PINN-predicted pressure fluctuations is poorest, with errors on the order of 100\%, which highlights the risk of weakly enforcing the governing equations while using the equations to estimate an unobserved quantity.
% , but nonetheless notable since it is only possible by virtue of enforcing the equations since the pressure is not observed.  

\begin{figure}
    \centering
    \begin{subfigure}{1.\linewidth}
    \centering
    \includegraphics[width=1.0\textwidth]{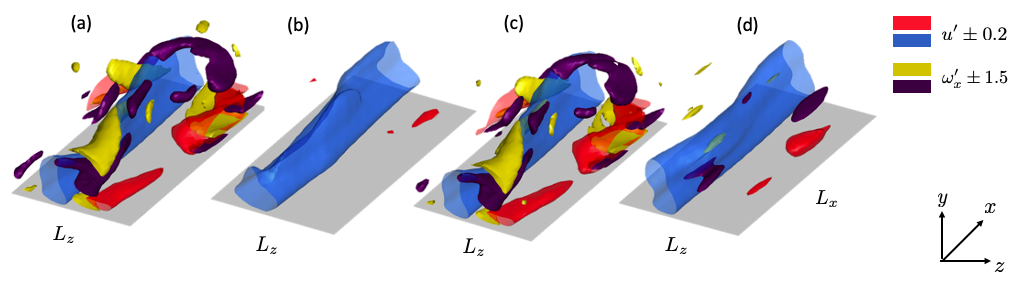}
    %\caption{$\epsilon_{u}$}
    \end{subfigure}
    \caption{Iso-surfaces of $u'$ and $\omega_x$ at $t=T$, when observations are separated by two Taylor microscales. 
    (a) Hidden truth; (b) spatio-temporal interpolation; (c) adjoint estimation case C-AJ; (d) PINN estimation case C-NN-S$_{\mathbf{W}}$.}
    \label{fig:3d contour 2Lambda}
\end{figure}

The above quantitative assessment of accuracy is interpreted visually in figure \ref{fig:3d contour 2Lambda}.  We plot iso-surfaces of the streamwise velocity and vorticity perturbation for case `C'.  Similar to the earlier visualization, the four panels correspond to the hidden true field, interpolation of the coarse observations, estimation using 4DVar and PINN reconstruction.   The interpolation field clearly highlights that the observations are scarce, incapable of capturing the velocity or vorticity structures.  The adjoint approach remain effective at estimating these turbulence structures, and demonstrates robustness to the sparsity of observations. The PINN reconstruction is, however, considerably less accurate in this case. 

Note that the network in case C-NN-S$_{\mathbf{W}}$ has the capacity to capture the fine-scale vortical structures in the flow, since it has identical size and architecture as case F-NN-S$_{\mathbf{W}}$ that was used for the finer observations (see e.g.~figure \ref{fig:3d contour 1Lambda}d).  As such, the notion that a larger network may be needed to improve accuracy is not applicable in the present case.  As for the training of the network, figure \ref{fig:2Lambda_error}f shows the history of the losses during training of C-NN-S$_{\mathbf{W}}$, which are all stagnated.  These losses are numerically larger than, but similar in orders of magnitude to those from training F-NN-S$_{\mathbf{W}}$ with fine observations (c.f.~figure \ref{fig:loss decay}). 
% The network is trained sufficiently with saturated observation, equation and boundary losses as shown in figure \ref{fig:2Lambda_error}f. 
Yet the present PINN-estimation contains little vortical structures compared to case `F'.
% in figure \ref{fig:3d contour 1Lambda}d. 

One potential reason for the difference in performance between 4DVar and PINN is their respective approaches to identifying a solution trajectory in state space.
In 4DVar, once an initial condition is estimated, the forward trajectory is unique and is fully determined by solving the discretized Navier-Stokes equations.  The optimization is thus only concerned with identifying the optimal initial condition that reproduces the observations.
The PINN solution trajectory is much more ambiguous because the equations are weakly enforced.  For every choice of the relative weighting between the measurement and physics loss ($\lambda_c$, $\lambda_u$, $\lambda_v$, $\lambda_w$ in equation \ref{eq:batch physics loss}), different trajectories become possible.  Consider for example the limiting behaviors when either the measurements or the equations weightings are relatively negligible; in the former a trivial solution exactly satisfies the governing equations and in the latter the network will simply interpolate the measurements with complete disregard to the physics.  We herein adopted the same strategy that was successful for the finer observations, but it is conceivable that a different weighting of the measurements and physics losses would produce better outcome. Our results should not be viewed as a verdict on the accuracy of PINN, but rather as a cautionary example and motivation for developing robust approaches that, for example, optimally weigh the different losses in order to identify the best solution trajectory. 

% \yd{One potential reason for the poor performance for PINN is the lack of deterministic causality.  The adjoint method searches for a deterministic trajectory that reproduces the data of all observation time. Through the strong constraints of physical equations, all of the unphysical trajectories that do not exactly satisfy equation are naturally ruled out. In PINN method, however, there is no deterministic trajectory defined by causality on time. The equations are enforced weakly, thus there could potentially be many solutions that both reproduce the observation data and weakly satisfy the governing equations. The value of these solutions away from observation locations are loosely controlled by the weak physical constraints, and are normally different from reference values.}

%The second reason lies on the  In adjoint method, the solution trajectory is searched through the information provided by the observations. All observations contribute to the search of the initial condition that reproduces this trajectory, thus the effect of observation is global on time and space. For PINN method, The observation at a certain spatio-temporal location do not contribute strongly to the inference of solution at earlier or later time due to the weakly enforced equations. The solution at certain point is only inferred from the observations near its location, which contain much less information compared to all observations.

%-------------------------------------
%   Influence of functional basis
%-------------------------------------
\subsection{Influence of PINN architecture}
\label{sec:results_basis}

The PINN architecture determines the function basis adopted in attempting to reproduce the observations and satisfy the governing equations.  As such, it is important to develop an appreciation for the impact of the network design on accuracy.  This point is amplified in the present context where the observations are generated by an independent direct numerical simulation algorithm that uses finite-volume discretization.  The discrepancy in the function basis that generated the measurements and adopted in PINN should not be viewed as a weakness of the present study, but rather as a surrogate to real scenarios where the measurements are obtained from independent physical experiments.

% \begin{figure}
%     \centering
%     \begin{subfigure}{.24\linewidth}
%     \centering
%     \includegraphics[width=1.0\textwidth]{Re180/NN_func_basis/error_u.pdf}
%     %\caption{$\epsilon_{u}$}
%     \end{subfigure}%
%     %
%     \begin{subfigure}{.24\linewidth}
%     \centering
%     \includegraphics[width=1.0\textwidth]{Re180/NN_func_basis/error_v.pdf}
%     %\caption{$\epsilon_{v}$}
%     \end{subfigure}%
%     %
%     \begin{subfigure}{.24\linewidth}
%     \centering
%     \includegraphics[width=1.0\textwidth]{Re180/NN_func_basis/error_w.pdf}
%     %\caption{$\epsilon_{w}$}
%     \end{subfigure}%
%     %
%     \begin{subfigure}{.24\linewidth}
%     \centering
%     \includegraphics[width=1.0\textwidth]{Re180/NN_func_basis/error_p.pdf}
%     %\caption{$\epsilon_{p}$}
%     \end{subfigure}
%     \caption{Estimation error of velocity and pressure fields for networks with different structures. (a-d) Panels show the $L^2$ error of reconstructed velocity and pressure fluctuations. \fullgray{}: F-NN-V, \dashedblack{}: F-NN-W, \dotted{}: F-NN-S$_{\mathbf{I}}$, : \chainred{}: F-NN-S$_{\mathbf{W}}$.}
%     \label{fig:NN structure error}
% \end{figure}

\begin{figure}
    \centering
    \begin{subfigure}{1.0\linewidth}
    \centering
    \includegraphics[width=1.0\textwidth]{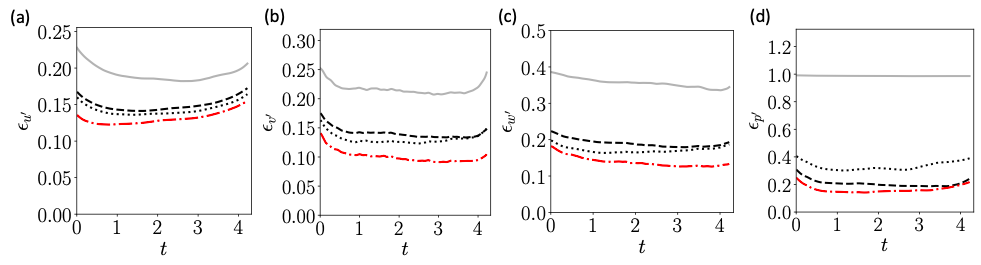}
    %\caption{$\epsilon_{u}$}
    \end{subfigure}%
    \caption{Estimation error of velocity and pressure fields for networks with different structures. (a-d) Panels show the $L^2$ error of reconstructed velocity and pressure fluctuations. \fullgray{}: F-NN-V, \dashedblack{}: F-NN-W, \dotted{}: F-NN-S$_{\mathbf{I}}$, : \chainred{}: F-NN-S$_{\mathbf{W}}$.}
    \label{fig:NN structure error}
\end{figure}

%TZ   One of the networks is F-NN-S$_{\mathbf{W}}$ which was adopted in all the previous results because it yields the best outcome.  This ResNet is comprised of 10 hidden layers each with 200 neurons, and is the benchmark for examining the performance of the other three.  

In figure \ref{fig:NN structure error}, we compare the performance of four PINN architectures. The discussion proceeds from the least to most accurate case, with the last one being F-NN-S$_{\mathbf{W}}$ which was adopted in all the previous results.  
The first network, `F-NN-V', is a vanilla, fully connected PINN that is relatively shallow with 4 hidden layers each comprised of 350 neurons.  Its estimation of the flow field yields the largest errors for all flow variables.  
The second network `F-NN-W' is designed to assess the effect of increasing the width of the network, by increasing the number of neurons per hidden layer to 500. 
In general, wider network have better capacity to learn complex functions, but are prone to overfitting. This concern is, however, mitigated in the PINNs framework because the equations loss is a natural regularization of the network functions away from the observation locations.  Indeed the results demonstrate that F-NN-W yields improved predictions, especially for the pressure. 

The remaining two PINNs are F-NN-S$_{\mathbf{I}}$ and F-NN-S$_{\mathbf{W}}$, which correspond to skip-connection architectures ResNet1 (\ref{eq:skip1}) and ResNet2 (\ref{eq:skip2}), respectively.
Both networks are comprised of 10 hidden layers each with 200 neurons.
ResNets mitigate the difficulty of training deep networks by adopting skip connections, thus facilitating efficient computation of the gradients flow through the networks \citep{he2016deep}.  
The results in figure \ref{fig:NN structure error} demonstrate that both F-NN-S$_{\mathbf{I}}$ and F-NN-S$_{\mathbf{W}}$ yield the most accurate estimation of the velocity and pressure fields.
It is important to note that the total number of trainable parameters in these networks is similar to the vanilla case F-NN-V, which has the worst performance.  The superior accuracy of the ResNet architectures in predicting the pressure, which is not observed, is entirely dependent on minimization of the equations loss.  These results underscore that the network structure is important for accurate realization of governing equations.  
%   Performance in the present test was, however, evaluated against the true solution which is not known in the case of field measurements for example.  In such cases, the design of a PINN should be based on theoretical estimation of its representation capability. % , which is a topic of continued research. 

%-----------------------------
%       Hard constraints
%-----------------------------
\subsection{PINN with embedded constraints}
\label{sec:results_constraints}
\begin{figure}
    \centering
    \begin{subfigure}{1.\linewidth}
    \centering
    \includegraphics[width=0.75\textwidth]{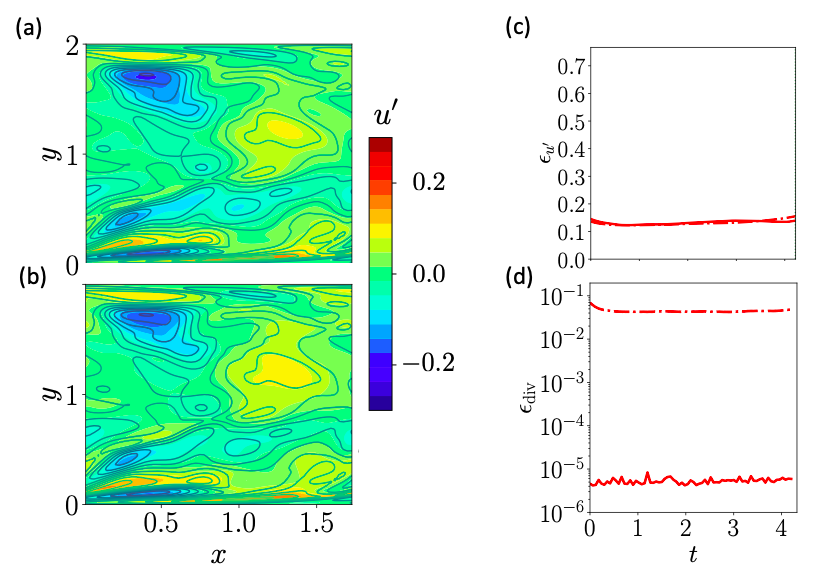}
    %\caption{$\epsilon_{u}$}
    \end{subfigure}
    \caption{Errors in the divergence-free condition of the estimated velocity fields. Contours of $u'$ from (a) F-NN-S$_{\mathbf{W}}$ and (b) F-NN-S$_{\mathbf{W}}$-EC. 
    (c) Estimation error in $u^{\prime}$ from (\chainred{}) F-NN-S$_{\mathbf{W}}$ and (\fullred) F-NN-S$_{\mathbf{W}}$-EC.
    (d) Volume-averaged error in divergence-free condition.}
    \label{fig:div free}
\end{figure}

%   In this section, 
An important difference between 4DVar and PINNs is the treatment of the equations and boundary conditions, which are enforced strongly in the former approach and only weakly in PINNs.  As such, important physics constraints, such as the requirement that the estimated velocity field is solenoidal, are only approximate in PINNs.  Some constraints can, however, be strongly enforced in PINNs with appropriate network design, as discussed in section \ref{sec:PINNalgorithm}.  

We embed the constraints for periodicity in the horizontal directions (equations \ref{eq:periodic BC}) and for a solenoidal velocity field (equation \ref{eq: vector potential}) into a ResNet1 that we designate F-NN-S$_{\mathbf{W}}$-EC.  Performance will be compared to the reference F-NN-S$_{\mathbf{W}}$, which has the same number of layers and neurons, and yielded the most accurate predictions using weak constraints. 
Contours of $u'$ for both cases are shown in figure \ref{fig:div free} along with their spatially averaged and normalized residual in the divergence constraint: 
\be
    \epsilon_{\mathrm{div}}=
    \frac{1}{\Omega} \int_{\Omega} 
    \frac{\vert\boldsymbol{\nabla} \cdot \boldsymbol{u}\vert}
        {\left(\left({\partial u}/{\partial x}\right)^{2}+\left({\partial v}/{\partial y}\right)^{2}+\left({\partial w}/{\partial z}\right)^{2}\right)^{1/2}} 
    \mathrm{~d} \boldsymbol{x}.
\ee
The contours of $u'$ in figure \ref{fig:div free} are visually similar to the truth for both networks, without and with the embedded divergence-free condition, and the errors are quantitatively similar (figure \ref{fig:div free}(c)). However, the accuracy of enforcing local mass conservation is appreciably different for the two networks (figure \ref{fig:div free}(d)).  Specifically, the accuracy of local mass conservation in case F-NN-S$_{\mathbf{W}}$ is limited, with minimum value on the order of $4\times10^{-2}$ which is undesirable in studies of incompressible turbulence.  This poor enforcement of the solenoidal condition is because the network attempts to minimize a multi-objective loss function that contains the divergence-free condition as just one of its objectives.  In contrast, network F-NN-S$_{\mathbf{W}}$-EC which adopted the embedded constraints has a residual on the order of $3\times 10^{-6}$. These results were obtained when single precision was adopted in the training and evaluation of the neural networks; higher precision would yield even lower values of $\epsilon_{\mathrm{div}}$, but would incur higher computational cost.

%YD Notes: Bullet points, at most three lines. Figure information, panels. 

% More specific on the bulletpoint

% List advantages and disadvantages 

%2.1 setup: the non-dimensionaliztion. Fix table locations. 
% Add a unified loss definition. 
% Loss go to the start of PINN sections
%Subsection about hard constraints, curl, sin-cos, 
% Caption of PINN schematics
% Results: Loss of Adjoint and PINNs. % Best results from adjoint & best results from neural networks. Both of them work. 
% Black pictures with shown pixcel. 
% Line truth, color contours
% Loss below 
% 
% Add LSE, 
% EnKF, between use equation and not equations 
% EnVar
% End of first paragraph with summary statement
% Add citations for introductions
% What people have used PINNs for. Adjoint 
% Wang & wang & zaki archive Hessian paper for continuous adjoint 
% Move numerical scheme to data generation
%discrete adjoint > continuous

% Schematic (a) adjoint (b) pinn

% optimization ,adjoint + PINN multiple curve, pre-multiply or different axis. Add the total loss

% combine 3.1 and 3.2
% contours u, v, w. For u, subtract the mean. 
% swap 3.2 and 3.3
% Truth, interpolation adjoint PINN order
% Combine 9 and 10 and put at first
% Combine 7 and 8
% move vortical structure to 3.2 for 1 lambda (or compute lambda_2)
%move 10 to before

%=============================
%       CONCLUSION
%=============================
\section{Conclusion}
\label{sec:conclusion} 

In this study, the state of turbulence in a minimal channel configuration was estimated from sparse measurements using adjoint variational data assimilation (4DVar) and physics-informed neural network (PINN).  The measurements were sub-sampled velocity data from an independent, fully resolved direct numerical simulation.  The flow estimation problem was formulated as a constrained optimization, where a loss function was defined in terms of the mismatch between the available flow observations and their prediction and the flow equations serve as the constraints.  
In adjoint method, a Lagrangian function combines the loss and the constraints, and its optimality condition yields the set of forward and adjoint equations that can be solved to evaluate the gradient of the loss with respect to the initial flow state.  Using this gradient, the loss is minimized when the optimal initial flow state is reached. 
In PINNs method, the physical equations and boundary conditions are enforced weakly, in $L^2$ sense.  In this form, they can also be regarded as regularization terms in the loss function.  The flow field is parameterized using the neural network.  By minimizing the full loss function, we obtain a network that reproduces the measurements at the observation locations and approximates the solution of the governing equations.  The estimated velocity and pressure fields are then obtained by sampling the trained network. The numerical results from both 4DVar and PINNs were evaluated in detail. Reconstruction from sparse observations with two resolutions were performed in order to assess the impact on the estimation accuracy and the robustness of the algorithms. 
%\yd{The relative performance of 4DVar and PINN were evaluated at two different Reynolds numbers.
%}
%\yd{The comparison of reconstruction accuracy from two methods is performed at different Reynolds number, }
For PINNs, different neural network structures were examined in order to demonstrate the influence of network basis on predictions. 

The comparative study highlighted a number of important points.  First, the major difference between 4Dvar and PINNs is the treatment of physical equations. While 4DVar strongly enforces the equations, PINNs adopts a weak $L^2$ form.  The accuracy of flow estimation in PINNs therefore depends on the relative weighting among observations, equations and boundary conditions losses, and these weights are often chosen empirically without rigorous theoretical guidance. 
%   In adjoint method, the equations and boundary conditions are strongly enforced, thus the optimization only focuses on the data observations. 
Nonetheless, our PINNs were effective in making accurate predictions of the flow state, even when the observations were acquired from an independent model with a different function basis.  Secondly, an important difference is that 4DVar enforces time causality of the flow state, exactly, while PINNs treat all instants in time as independent samples.  In addition to the conceptual nature of this difference, one practical implication is the accuracy of the predicted fields.  In 4DVar, the maximum accuracy is achieved at the end of the observation horizon, while the accuracy of PINNs predictions are nearly constant over the observation time (unless the measurements are weighted in the loss). The difference is relevant to the third point, which is the accuracy of forecasts.  While 4DVar is ideally suited for forecasts, PINNs are notoriously inaccurate beyond their training horizon.  Therefore either additional training is required for forecasts or their estimated flow fields must be evolved by another forward model (e.g.\,conventional CFD algorithms or evolutional deep neural network \citep{du2021ednn}).

Another important consideration relates to design of the algorithms.  Adjoint-variational methods are based on numerical discretizations, where the convergence rate and error estimation are rigorously established by numerical analysis. In contrast, the theoretical foundation for convergence of neural networks in approximation of functions remains subject of active research.  This accuracy of PINN, which depends on its representation capability, is therefore difficult to anticipate for a given network architecture and size.  

There are several questions that should be addressed by future efforts.  Firstly, the computational costs of 4DVar and PINN should be compared.  For the present study, the 4DVar algorithm was impelemnted on CPU while the PINN was trained using GPU, and therefore comparison of their wall-clock computational time is not meaningful.  A comparison based on floating-point operations or power consumption, for the same level of accuracy, would be beneficial.  Secondly, the Reynolds numbers considered in this study are moderate and the spatio-temporal domain was limited in size. Future comparisons should target higher Reynolds numbers and larger domains.  Finally, the present study used an independent direct numerical simulation as a surrogate for generating experimental measurements. Future efforts should consider estimation of the flow state using measurements from experiments, where some of the measurements can be retained as independent data for a blind evaluation of accuracy.

%========================================
%       ACKNOWLEDGEMENTS
%========================================

\par\bigskip
\noindent
\textbf{Acknowledgements.} 
The authors acknowledge financial support from the Office of Naval Research (N00014-20-1-2715,  N00014-21-1-2375). Computational resources were provided by the Maryland Advanced Research Computing Center (MARCC).

\par\bigskip
\noindent
\textbf{Declaration of interests.} 
The authors report no conflict of interest.

\appendix

\section{Forward and adjoint boundary variables}
For completeness of the adjoint-variational formulation, a detailed description of forward and adjoint variables $\boldsymbol{\beta}$ and $\boldsymbol{\beta}^{\dag}$ is provided. The forward boundary variable $\boldsymbol{\beta}$ is defined as a column vector with $12$ entries, comprised of the velocities and pressure on the $6$ faces of the computational domain,
 \be
 	\boldsymbol{\beta} = 
 	\begin{pmatrix}
 	\boldsymbol{u}(x=0,y,z,t)\\
 	\boldsymbol{u}(x=L_x,y,z,t)\\
 	\boldsymbol{u}(x,y=0,z,t)\\
 	\boldsymbol{u}(x,y=L_y,z,t)\\
 	\boldsymbol{u}(x,y,z=0,t)\\
 	\boldsymbol{u}(x,y,z=L_z,t)\\
 	p(x=0,y,z,t)\\
 	p(x=L_x,y,z,t)\\
 	p(x,y=0,z,t)\\
 	p(x,y=L_y,z,t)\\
 	p(x,y,z=0,t)\\
 	p(x,y,z=L_z,t)\\
 	\end{pmatrix}. 
 \ee
The forward boundary constraint $\mathscr{B}[\boldsymbol{u},p]=0$ is a column vector of equations containing in total $8$ boundary conditions: periodicity of $\boldsymbol{u}$ and $p$ in the horizontal  $x$ and $z$ directions; no-slip boundary conditions for $\boldsymbol{u}$ and homogeneous Neumann boundary conditions for $p$ on the walls. These conditions are given below: 
 \be
 	\mathscr{B}[\boldsymbol{u},p] = 
 	\begin{pmatrix}
 	\boldsymbol{u}(x=0,y,z,t)-\boldsymbol{u}(x=L_x,y,z,t)\\
 	\boldsymbol{u}(x,y=0,z,t)\\
 	\boldsymbol{u}(x,y=L_y,z,t)\\
 	\boldsymbol{u}(x,y,z=0,t)-\boldsymbol{u}(x,y,z=L_z,t)\\
 	p(x=0,y,z,t)-p(x=L_x,y,z,t)\\
 	\mathrm{d}p/\mathrm{d}y(x,y=0,z,t)\\
 	\mathrm{d}p/\mathrm{d}y(x,y=L_y,z,t)\\
 	p(x,y,z=0,t)-p(x,y,z=L_z,t)\\
 	\end{pmatrix}. 
 \ee

The adjoint boundary variable $\boldsymbol{\beta}^{\dag}$ is the Lagrange multiplier of the forward boundary constraint $\mathscr{B}[\boldsymbol{u},p] = 0$, and has the same dimension as the boundary condition constraints $\mathscr{B}[\boldsymbol{u},p]$.  The general form of $\boldsymbol{\beta}^{\dag}$ is given below:  
\be
 	\boldsymbol{\beta}^{\dag} = 
 	\begin{pmatrix}
 	\boldsymbol{\beta}^{\dag}_{\boldsymbol{u},x}(y,z,t)\\
 	\boldsymbol{\beta}^{\dag}_{\boldsymbol{u},0}(x,z,t)\\
 	\boldsymbol{\beta}^{\dag}_{\boldsymbol{u},L_y}(x,z,t)\\
 	\boldsymbol{\beta}^{\dag}_{\boldsymbol{u},z}(x,y,t)\\
 	\beta^{\dag}_{p,x}(y,z,t)\\
 	\beta^{\dag}_{p,0}(x,z,t)\\
 	\beta^{\dag}_{p,L_y}(x,z,t)\\
 	\beta^{\dag}_{p,z}(x,z,t)\\
 	\end{pmatrix}. 
 \ee
 
\section{Comparison of 4DVar and PINNs at $\mathrm{Re}_{\tau}=392$}\label{sec:appendix B}

In this section, we examine the influence of a higher Reynolds number of the accuracy of the state estimation using both 4DVar and PINN.  The comparison is performed at bulk Reynolds number $\mathrm{Re}=6875$, or equivalently $\mathrm{Re}_{\tau}=392$.
The reference simulation for collecting observations is of flow in a minimal channel unit, where the parameters are listed in table \ref{table:channel setup Re392}. 
The observation data are generated by sub-sampling the velocity fields of the reference simulation in space and time.  The sub-sampling parameters (see table \ref{table:channel setup Re392}) are consistent with what is regarded as fine measurement resolution (designation `F-392'), where samples are separated by one Taylor microscale.  The duration of assimilation window is $T^{+} = 50$, which is the same $T^{+}$ adopted in the main text.

Using the velocity observations, the adjoint (`F-AJ-392') and PINN (`F-NN-S$_{\mathbf{W}}$-392') methods are utilized to estimate the full spatio-temporal velocity and pressure fields.  The computational domain and grid used in the adjoint assimilation `F-AJ-392' share the same parameters as the reference simulations. For PINN, the parameters of the network for case `F-NN-S$_{\mathbf{W}}$-392' are provided in table \ref{table:NN Re392}.  The number of network parameters to be optimized, $\mathrm{dim}(\boldsymbol{\theta})$, is larger than at the lower Reynolds number.  Nonetheless, this value remains much smaller than the dimension of the initial state, $\mathrm{dim}(\boldsymbol{u}_0)=11 \times 10^6$.

\begin{figure}
    \centering
    \includegraphics[width=1.0\textwidth]{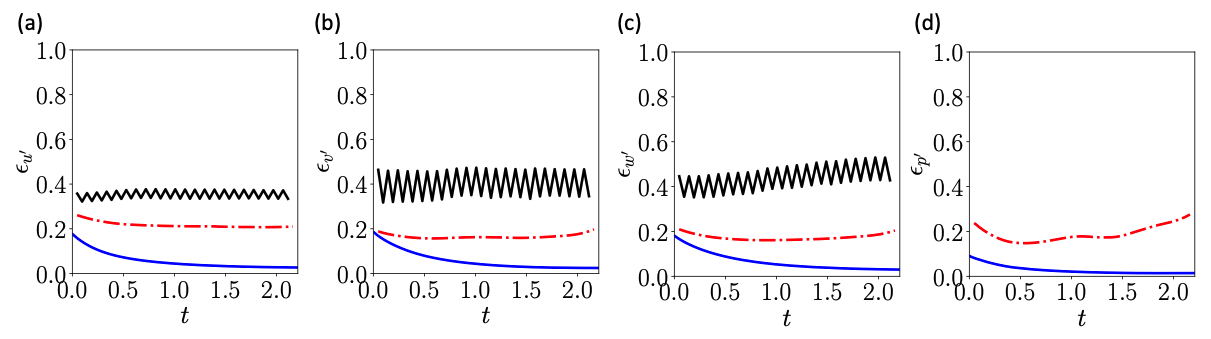}
    \caption{Estimation errors of minimal channel flow at $\mathrm{Re}_{\tau}=392$ when observations are separated by one Taylor microscale. (a-d) $L^2$ error of estimated perturbation velocities and pressure. \fullblack{}: Spatio-temporal interpolation,  \full{}: adjoint estimation case F-AJ-392, \chainred{}: PINNs estimation case F-NN-S$_{\mathbf{W}}$-392.  }
    \label{fig:Re392 1Lambda_error}
\end{figure}

\begin{figure}
    \centering
    \begin{subfigure}{1.0\linewidth}
    \centering
    \includegraphics[width=1.0\textwidth]{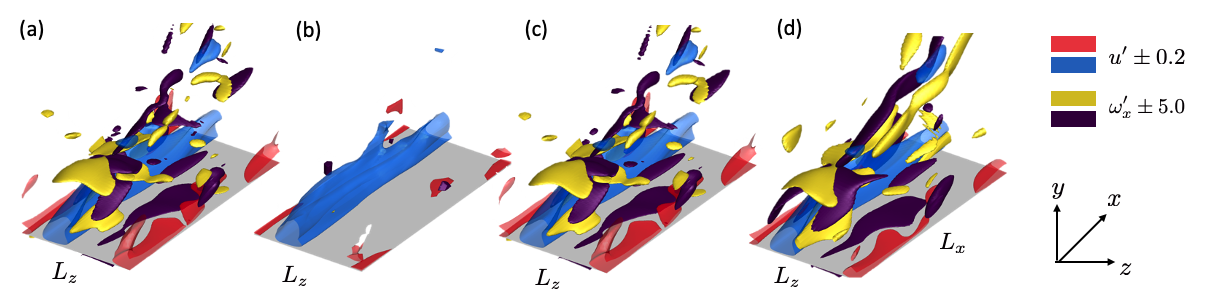}
    \end{subfigure}
    \caption{Iso-surfaces of $u'$ and $\omega'_x$ at $t=T$, when observations are separated by one Taylor microscale. (a) Hidden truth; (b) spatio-temporal interpolation; (c) 4DVar estimation case F-AJ-392; (d) PINN estimation case F-NN-S$_{\mathbf{W}}$-392.}
    \label{fig:Re392 3d contour 1Lambda}
\end{figure}

The root-mean-square errors (\ref{eq: rms error}) in the velocity perturbations and pressure 
are reported in figure \ref{fig:Re392 1Lambda_error}.  Each panel compares the errors when the field is estimated using mere interpolation of the observations, 4DVar and PINN.  Qualitatively the results are similar to those obtained at lower Reynolds number (c.f. figure \ref{fig:1Lambda_error}). Specifically, the estimated states from 4DVar and PINN are more accurate than from interpolation, and the predictions by 4DVar are the most accurate.  The final flow state, at $t=T$, is visualized in figure \ref{fig:Re392 3d contour 1Lambda}.  The isosurfaces mark two levels of streamwise velocity perturbations and two levels of streamwise vorticity.  The comparison of the true field, interpolation of observations, 4DVar and PINN again underscores the capacity of the last two techniques to estimate vorticity structures that are absent in the observation data. The 4DVar estimation is again most accurate when compared to the true state.

% 	\begin{table}
% 		\centering
% 		\begin{tabular}{ccccc}
% 			Domain size & Grid numbers&  Grid resolution  & Observations \\
			
% 			\begin{tabular}{c}
% 				\hline 
% 				($L_x^{+}$, $L_y^{+}$, $L_z^{+}$) \\ \hline 
% 			(314, 784, 157)   \\
% 			\end{tabular} & 
% 			\begin{tabular}{c}
% 				\hline
% 				($N_x$, $N_y$, $N_z$) \\ \hline 
% 			%\rowcolor{blue!20}	
% 			(65, 897, 65)  \\
% 			%\tz{64}    & \tz{896}    & \tz{64}
% 			\end{tabular} & 
% 			\begin{tabular}{c}
% 				\hline
% 				($\Delta x^+$, $\Delta y^+$, $\Delta z^+$) \\ \hline 
% 				%\rowcolor{blue!20}	
% 				(4.90, 0.40 -- 1.40, 2.45)  \\
% 			\end{tabular} &
% 			\begin{tabular}{c}
% 				\hline
% 				($\Delta M_x$, $\Delta M_y$, $\Delta M_z$, $\Delta M_t$)  \\ \hline 
% 		      (6, 32, 4, 32)          \\
% 		    %\tz{6875}            &   \tz{392}
% 			\end{tabular}\\
% 		\end{tabular}
% 		\caption{\tz{Domain size and grid resolution for reference DNS, and the sub-sampling for extracting observations.  The adjoint data assimilation (case F-AJ-392) adopts the same grid as the reference simulation.}}
% 		\label{table:channel setup Re392}
% 	\end{table}

	\begin{table}
		\centering
		\begin{tabular}{ccc}
			Domain size & Grid numbers&  Grid resolution  \\
			
			\begin{tabular}{ccc}
				\hline 
				$L_x^{+}$ & $L_y^{+}$ & $L_z^{+}$ \\ \hline 
			314 & 784 & 157   \\
			\end{tabular} & 
			\begin{tabular}{ccc}
				\hline
				$N_x$ & $N_y$ & $N_z$ \\ \hline 
			%\rowcolor{blue!20}	
			65 & 897 & 65  \\
			%\tz{64}    & \tz{896}    & \tz{64}
			\end{tabular} & 
			\begin{tabular}{cccc}
				\hline
				$\Delta x^+$& $\Delta y^+_{\mathrm{min}}$& $\Delta y^+_{\mathrm{max}}$&
				$\Delta z^+$ \\ \hline 
				%\rowcolor{blue!20}	
				4.90 & 0.40 & 1.40 & 2.45  \\
			\end{tabular} \\
		\end{tabular}
				\begin{tabular}{c}
				Observations \\
		  	\begin{tabular}{cccc}
				\hline
				$\Delta M_x$& $\Delta M_y$& $\Delta M_z$& $\Delta M_t$  \\ \hline 
		      6 & 32 & 4 & 32          \\
		    %\tz{6875}            &   \tz{392}
			\end{tabular}\\
			\end{tabular}
		\caption{Domain size and grid resolution for reference DNS, and the sub-sampling for extracting observations.  The adjoint data assimilation (case F-AJ-392) adopts the same grid as the reference simulation.}
		\label{table:channel setup Re392}
	\end{table}

\begin{table}
	\centering
		\begin{tabular}{l c c c c c c c c }
		    \hline
	 	     Case & $D$ & $n_{L}$ & structure & $\mathrm{dim}(\boldsymbol{\theta})$& section  \\
			\hline 
			F-NN-S$_{\mathbf{W}}$-392 & 15 & 200 & ResNet2 & $6.85\times10^{5}$ & \S \ref{sec:appendix B}  \\
			\hline 
	\end{tabular}
	\caption{Parameters of PINNs for the flow estimation at $\mathrm{Re}=392$.}
	\label{table:NN Re392}
\end{table}

\newpage
\bibliography{reference}

\begin{thebibliography}{47}
\providecommand{\natexlab}[1]{#1}
\providecommand{\url}[1]{\texttt{#1}}
\expandafter\ifx\csname urlstyle\endcsname\relax
  \providecommand{\doi}[1]{doi: #1}\else
  \providecommand{\doi}{doi: \begingroup \urlstyle{rm}\Url}\fi

\bibitem[Evensen(1994)]{Evensen1994EnKF}
Geir Evensen.
\newblock Sequential data assimilation with a nonlinear quasi-geostrophic model
  using monte carlo methods to forecast error statistics.
\newblock \emph{J. Geophys. Res. Oceans}, 99\penalty0 (C5):\penalty0
  10143--10162, 1994.

\bibitem[Suzuki and Hasegawa(2017)]{Suzuki2017}
Takao Suzuki and Yosuke Hasegawa.
\newblock Estimation of turbulent channel flow at {$Re_{\tau}=100$} based on
  the wall measurement using a simple sequential approach.
\newblock \emph{J.~Fluid Mech.}, 830:\penalty0 760--796, 2017.

\bibitem[Zaki and Wang(2021)]{zaki2021prf}
Tamer~A. Zaki and Mengze Wang.
\newblock From limited observations to the state of turbulence: Fundamental
  difficulties of flow reconstruction.
\newblock \emph{Phys. Rev. Fluids}, 6:\penalty0 100501, Oct 2021.

\bibitem[Buchta and Zaki(2021)]{Buchta2021envar}
David~A. Buchta and Tamer~A. Zaki.
\newblock Observation-infused simulations of high-speed boundary-layer
  transition.
\newblock \emph{J.~Fluid Mech.}, 916:\penalty0 A44, 2021.

\bibitem[Buchta et~al.(2022)Buchta, Laurence, and Zaki]{Buchta2022}
David~A. Buchta, Stuart~J. Laurence, and Tamer~A. Zaki.
\newblock Assimilation of wall-pressure measurements in high-speed flow over a
  cone.
\newblock \emph{Journal of Fluid Mechanics}, 947:\penalty0 R2, 2022.

\bibitem[Raissi et~al.(2019)Raissi, Perdikaris, and
  Karniadakis]{raissi2019physics}
Maziar Raissi, Paris Perdikaris, and George~E Karniadakis.
\newblock Physics-informed neural networks: A deep learning framework for
  solving forward and inverse problems involving nonlinear partial differential
  equations.
\newblock \emph{Journal of Computational physics}, 378:\penalty0 686--707,
  2019.

\bibitem[Adrian and Moin(1988)]{Adrian1988LSE}
Ronald~J Adrian and Parviz Moin.
\newblock Stochastic estimation of organized turbulent structure: homogeneous
  shear flow.
\newblock \emph{J.~Fluid Mech.}, 190:\penalty0 531--559, 1988.

\bibitem[Encinar and Jim{\'e}nez(2019)]{Jimenez2019LSE}
Miguel~P Encinar and Javier Jim{\'e}nez.
\newblock Logarithmic-layer turbulence: {A} view from the wall.
\newblock \emph{Phys. Rev. Fluids}, 4\penalty0 (11):\penalty0 114603, 2019.

\bibitem[Chevalier et~al.(2006)Chevalier, Hœpffner, Bewley, and
  Henningson]{Bewley_part2}
Mattias Chevalier, Jérôme Hœpffner, Thomas~R. Bewley, and Dan~S. Henningson.
\newblock State estimation in wall-bounded flow systems. part 2. turbulent
  flows.
\newblock \emph{J.~Fluid Mech.}, 552:\penalty0 167–187, 2006.

\bibitem[Suzuki(2012)]{Suzuki2012}
Takao Suzuki.
\newblock Reduced-order kalman-filtered hybrid simulation combining particle
  tracking velocimetry and direct numerical simulation.
\newblock \emph{J.~Fluid Mech.}, 709:\penalty0 249–288, 2012.

\bibitem[Colburn et~al.(2011)Colburn, Cessna, and Bewley]{colburn2011state}
CH~Colburn, JB~Cessna, and TR~Bewley.
\newblock State estimation in wall-bounded flow systems. part 3. the ensemble
  kalman filter.
\newblock \emph{Journal of Fluid Mechanics}, 682:\penalty0 289--303, 2011.

\bibitem[Fukami et~al.(2019)Fukami, Fukagata, and Taira]{fukami2019super}
Kai Fukami, Koji Fukagata, and Kunihiko Taira.
\newblock Super-resolution reconstruction of turbulent flows with machine
  learning.
\newblock \emph{Journal of Fluid Mechanics}, 870:\penalty0 106--120, 2019.

\bibitem[Gundersen et~al.(2021)Gundersen, Oleynik, Blaser, and
  Alendal]{gundersen2021semi}
Kristian Gundersen, Anna Oleynik, Nello Blaser, and Guttorm Alendal.
\newblock Semi-conditional variational auto-encoder for flow reconstruction and
  uncertainty quantification from limited observations.
\newblock \emph{Physics of Fluids}, 33\penalty0 (1):\penalty0 017119, 2021.

\bibitem[Dimet and Talagrand(1986)]{Dimet1986_4dvar}
FranÇois-Xavier~Le Dimet and Olivier Talagrand.
\newblock Variational algorithms for analysis and assimilation of
  meteorological observations: theoretical aspects.
\newblock \emph{Tellus A: Dyn. Meteorol. Oceanogr.}, 38\penalty0 (2):\penalty0
  97--110, 1986.

\bibitem[Li et~al.(2020)Li, Zhang, Dong, and Abdullah]{Li2020}
Yi~Li, Jianlei Zhang, Gang Dong, and Naseer~S Abdullah.
\newblock Small-scale reconstruction in three-dimensional kolmogorov flows
  using four-dimensional variational data assimilation.
\newblock \emph{J.~Fluid Mech.}, 885:\penalty0 A9, 2020.

\bibitem[Wang et~al.(2019{\natexlab{a}})Wang, Hasegawa, and
  Zaki]{Wang_hasegawa_zaki_2019}
Qi~Wang, Yosuke Hasegawa, and Tamer~A. Zaki.
\newblock Spatial reconstruction of steady scalar sources from remote
  measurements in turbulent flow.
\newblock \emph{J.~Fluid Mech.}, 870:\penalty0 316–352, 2019{\natexlab{a}}.

\bibitem[Wang et~al.(2019{\natexlab{b}})Wang, Wang, and Zaki]{wang2019discrete}
Mengze Wang, Qi~Wang, and Tamer~A Zaki.
\newblock Discrete adjoint of fractional-step incompressible {N}avier-{S}tokes
  solver in curvilinear coordinates and application to data assimilation.
\newblock \emph{Journal of Computational Physics}, 396:\penalty0 427--450,
  2019{\natexlab{b}}.

\bibitem[Bewley and Protas(2004)]{bewley2004skin}
Thomas~R Bewley and Bartosz Protas.
\newblock Skin friction and pressure: the “footprints” of turbulence.
\newblock \emph{Physica D: Nonlinear Phenomena}, 196\penalty0 (1-2):\penalty0
  28--44, 2004.

\bibitem[Foures et~al.(2014)Foures, Dovetta, Sipp, and Schmid]{Foures2014}
D.~P.~G. Foures, N.~Dovetta, D.~Sipp, and P.~J. Schmid.
\newblock A data-assimilation method for {Reynolds-averaged
  Navier-Stokes-driven} mean flow reconstruction.
\newblock \emph{J.~Fluid Mech.}, 759:\penalty0 404--431, 2014.

\bibitem[Wang and Zaki(2021)]{wang2021state}
Mengze Wang and Tamer~A Zaki.
\newblock State estimation in turbulent channel flow from limited observations.
\newblock \emph{Journal of Fluid Mechanics}, 917:\penalty0 A9, 2021.

\bibitem[Wang et~al.(2022)Wang, Wang, and Zaki]{wang2022hessian}
Qi~Wang, Mengze Wang, and Tamer~A. Zaki.
\newblock What is observable from wall data in turbulent channel flow?
\newblock \emph{Journal of Fluid Mechanics}, 941:\penalty0 A48, 2022.

\bibitem[Mons et~al.(2019)Mons, Wang, and Zaki]{mons2019kriging}
Vincent Mons, Qi~Wang, and Tamer~A Zaki.
\newblock Kriging-enhanced ensemble variational data assimilation for
  scalar-source identification in turbulent environments.
\newblock \emph{J.~Comput. Phys.}, 398:\penalty0 108856, 2019.

\bibitem[Mons et~al.(2021)Mons, Du, and Zaki]{mons2021les}
Vincent Mons, Yifan Du, and Tamer~A. Zaki.
\newblock Ensemble-variational assimilation of statistical data in large-eddy
  simulation.
\newblock \emph{Phys. Rev. Fluids}, 6:\penalty0 104607, Oct 2021.

\bibitem[Mons et~al.(2016)Mons, Chassaing, Gomez, and
  Sagaut]{mons2016reconstruction}
Vincent Mons, J-C Chassaing, Thomas Gomez, and Pierre Sagaut.
\newblock Reconstruction of unsteady viscous flows using data assimilation
  schemes.
\newblock \emph{Journal of Computational Physics}, 316:\penalty0 255--280,
  2016.

\bibitem[Mao et~al.(2020)Mao, Jagtap, and Karniadakis]{mao2020physics}
Zhiping Mao, Ameya~D Jagtap, and George~Em Karniadakis.
\newblock Physics-informed neural networks for high-speed flows.
\newblock \emph{Computer Methods in Applied Mechanics and Engineering},
  360:\penalty0 112789, 2020.

\bibitem[Lou et~al.(2021)Lou, Meng, and Karniadakis]{lou2021physics}
Qin Lou, Xuhui Meng, and George~Em Karniadakis.
\newblock Physics-informed neural networks for solving forward and inverse flow
  problems via the boltzmann-bgk formulation.
\newblock \emph{Journal of Computational Physics}, 447:\penalty0 110676, 2021.

\bibitem[Mao et~al.(2021)Mao, Lu, Marxen, Zaki, and Karniadakis]{mao2021jcp}
Zhiping Mao, Lu~Lu, Olaf Marxen, Tamer~A. Zaki, and George~Em Karniadakis.
\newblock Deep{M}\&{M}net for hypersonics: Predicting the coupled flow and
  finite-rate chemistry behind a normal shock using neural-network
  approximation of operators.
\newblock \emph{Journal of Computational Physics}, 447:\penalty0 110698, 2021.
\newblock ISSN 0021-9991.

\bibitem[Cai et~al.(2021)Cai, Wang, Lu, Zaki, and Karniadakis]{cai2021jcp}
Shengze Cai, Zhicheng Wang, Lu~Lu, Tamer~A. Zaki, and George~Em Karniadakis.
\newblock Deep{M}\&{M}net: Inferring the electroconvection multiphysics fields
  based on operator approximation by neural networks.
\newblock \emph{Journal of Computational Physics}, 436:\penalty0 110296, 2021.
\newblock ISSN 0021-9991.

\bibitem[Di~Leoni et~al.(2021)Di~Leoni, Lu, Meneveau, Karniadakis, and
  Zaki]{PCDL2021DeepONet}
P~Clark Di~Leoni, Lu~Lu, Charles Meneveau, George Karniadakis, and Tamer~A
  Zaki.
\newblock Deeponet prediction of linear instability waves in high-speed
  boundary layers.
\newblock \emph{arXiv preprint arXiv:2105.08697}, 2021.

\bibitem[Mowlavi and Nabi(2021)]{mowlavi2021optimal}
Saviz Mowlavi and Saleh Nabi.
\newblock Optimal control of pdes using physics-informed neural networks.
\newblock \emph{arXiv preprint arXiv:2111.09880}, 2021.

\bibitem[Yang et~al.(2021)Yang, Meng, and Karniadakis]{yang2021b}
Liu Yang, Xuhui Meng, and George~Em Karniadakis.
\newblock B-pinns: Bayesian physics-informed neural networks for forward and
  inverse pde problems with noisy data.
\newblock \emph{Journal of Computational Physics}, 425:\penalty0 109913, 2021.

\bibitem[Jim{\'e}nez and Moin(1991)]{jimenez1991minimal}
Javier Jim{\'e}nez and Parviz Moin.
\newblock The minimal flow unit in near-wall turbulence.
\newblock \emph{Journal of Fluid Mechanics}, 225:\penalty0 213--240, 1991.

\bibitem[Podvin and Lumley(1998)]{podvin1998low}
B{\'e}reng{\`e}re Podvin and John Lumley.
\newblock A low-dimensional approach for the minimal flow unit.
\newblock \emph{Journal of Fluid Mechanics}, 362:\penalty0 121--155, 1998.

\bibitem[Flores and Jim{\'e}nez(2010)]{flores2010hierarchy}
Oscar Flores and Javier Jim{\'e}nez.
\newblock Hierarchy of minimal flow units in the logarithmic layer.
\newblock \emph{Physics of Fluids}, 22\penalty0 (7):\penalty0 071704, 2010.

\bibitem[Jelly et~al.(2014)Jelly, Jung, and Zaki]{Jelly2014}
TO~Jelly, SY~Jung, and TA~Zaki.
\newblock Turbulence and skin friction modification in channel flow with
  streamwise-aligned superhydrophobic surface texture.
\newblock \emph{Phys. Fluids}, 26\penalty0 (9):\penalty0 095102, 2014.

\bibitem[You and Zaki(2019)]{you_zaki_2019}
Jiho You and Tamer~A. Zaki.
\newblock Conditional statistics and flow structures in turbulent boundary
  layers buffeted by free-stream disturbances.
\newblock \emph{Journal of Fluid Mechanics}, 866:\penalty0 526–566, 2019.

\bibitem[Rosenfeld et~al.(1991)Rosenfeld, Kwak, and Vinokur]{Rosenfeld1991}
M.~Rosenfeld, D.~Kwak, and M.~Vinokur.
\newblock A fractional step solution method for the unsteady incompressible
  {Navier-Stokes} equations in generalized coordinate systems.
\newblock \emph{J.~Comput. Phys.}, 94:\penalty0 102--137, 1991.

\bibitem[Wang and Zaki(2022)]{wang2022synch}
Mengze Wang and Tamer~A. Zaki.
\newblock Synchronization of turbulence in channel flow.
\newblock \emph{Journal of Fluid Mechanics}, 943:\penalty0 A4, 2022.

\bibitem[Nikitin(2018)]{Nikitin2018}
N.~Nikitin.
\newblock Characteristics of the leading {Lyapunov} vector in a turbulent
  channel flow.
\newblock \emph{J.~Fluid Mech.}, 849:\penalty0 942--967, 2018.

\bibitem[Nocedal(1980)]{LBFGS}
J.~Nocedal.
\newblock Updating {quasi-Newton} matrices with limited storage.
\newblock \emph{Math. Comput.}, 35\penalty0 (151):\penalty0 773--782, 1980.

\bibitem[Wang et~al.(2020)Wang, Teng, and Perdikaris]{wang2020understanding}
Sifan Wang, Yujun Teng, and Paris Perdikaris.
\newblock Understanding and mitigating gradient pathologies in physics-informed
  neural networks.
\newblock \emph{arXiv preprint arXiv:2001.04536}, 2020.

\bibitem[Poole et~al.(2016)Poole, Lahiri, Raghu, Sohl-Dickstein, and
  Ganguli]{poole2016exponential}
Ben Poole, Subhaneil Lahiri, Maithra Raghu, Jascha Sohl-Dickstein, and Surya
  Ganguli.
\newblock Exponential expressivity in deep neural networks through transient
  chaos.
\newblock \emph{Advances in neural information processing systems}, 29, 2016.

\bibitem[Li et~al.(2018)Li, Xu, Taylor, Studer, and
  Goldstein]{li2018visualizing}
Hao Li, Zheng Xu, Gavin Taylor, Christoph Studer, and Tom Goldstein.
\newblock Visualizing the loss landscape of neural nets.
\newblock \emph{Advances in neural information processing systems}, 31, 2018.

\bibitem[Veit et~al.(2016)Veit, Wilber, and Belongie]{veit2016residual}
Andreas Veit, Michael~J Wilber, and Serge Belongie.
\newblock Residual networks behave like ensembles of relatively shallow
  networks.
\newblock \emph{Advances in neural information processing systems}, 29, 2016.

\bibitem[He et~al.(2016)He, Zhang, Ren, and Sun]{he2016deep}
Kaiming He, Xiangyu Zhang, Shaoqing Ren, and Jian Sun.
\newblock Deep residual learning for image recognition.
\newblock In \emph{Proceedings of the IEEE conference on computer vision and
  pattern recognition}, pages 770--778, 2016.

\bibitem[Yazdani et~al.(2020)Yazdani, Lu, Raissi, and
  Karniadakis]{yazdani2020systems}
Alireza Yazdani, Lu~Lu, Maziar Raissi, and George~Em Karniadakis.
\newblock Systems biology informed deep learning for inferring parameters and
  hidden dynamics.
\newblock \emph{PLoS computational biology}, 16\penalty0 (11):\penalty0
  e1007575, 2020.

\bibitem[Du and Zaki(2021)]{du2021ednn}
Yifan Du and Tamer~A. Zaki.
\newblock Evolutional deep neural network.
\newblock \emph{Phys. Rev. E}, 104:\penalty0 045303, 2021.

\end{thebibliography}
\end{document}